\documentclass[a4paper,10pt]{article}
\usepackage{fullpage}
\usepackage{cite}
\usepackage{amsmath}
\usepackage{amssymb}
\usepackage{amsfonts}
\usepackage{graphicx}
\usepackage{caption}
\usepackage{subcaption}
\usepackage{float}
\usepackage[usenames,dvipsnames]{xcolor}
\usepackage{tikz}
\usepackage{pstricks}
\usepackage{authblk}
\usepackage{setspace}
\usepackage[pagebackref=true]{hyperref}
\hypersetup{
	colorlinks=true,
	linkcolor=blue,
	filecolor=Violet,     
	urlcolor=RubineRed,
	citecolor=red
}

\begin{document}
	
	
	
	
	
	\date{}
	
	\author{Pritam Banerjee \thanks{\href{mailto:bpritam@iitk.ac.in}{bpritam@iitk.ac.in}}}
	
	\author{Suvankar Paul \thanks{\href{mailto:svnkr@iitk.ac.in}{svnkr@iitk.ac.in}}}
	
	\author{Tapobrata Sarkar \thanks{\href{mailto:tapo@iitk.ac.in}{tapo@iitk.ac.in}}}
	
	\affil{Department of Physics,\\  Indian Institute of Technology,  \\ Kanpur 208016, India}
	
	\title{ On Strong Gravitational Lensing in Rotating Galactic Space-times}
	
	\maketitle

		\begin{abstract}
			\noindent
			We study strong gravitational lensing in rotating space-times which can be thought of as realistic galactic models in General Relativity. 
			To this end, using the Newman-Janis algorithm, we first obtain a rotating version of a static galactic metric advanced by
			Bharadwaj and Kar. This is matched with a generic external Kerr solution without a slowly-rotating approximation. Next, we construct a 
			rotating version of Perlick's Bertrand
			space-times and establish this as a new rotating solution of Einstein's equations where each spatial radius admits a stable circular 
			orbit. A rotating generalization 
			of a strong lensing formalism due to Perlick is then applied to the rotating Bharadwaj-Kar metric and the differences with lensing
			in the purely Kerr background (for a Kerr black hole at the galactic center) is pointed out. Then, arguing that a rotating Bertrand 
			space-time might be a 
			realistic galactic model away from the galactic center, strong lensing is studied in this 
			situation where the central singularity is naked. Possible observational signatures are
			elaborated upon, which might be useful in distinguishing black holes from naked singularities. 
	\end{abstract}

\section{Introduction and Motivation}

The phenomenon of gravitational lensing has been widely studied in the literature, and continues to be a topic of great interest,
as it promises to further our understanding of dark matter at galactic scales.\footnote{The literature on the topic is huge, and
for a small sampling of both theoretical and observational aspects, we refer the reader to \cite{Review1},\cite{Review2},\cite{Review3},
\cite{Review4},\cite{Review5},\cite{Review6},\cite{Review7},\cite{Review8},\cite{Review9}.} Several methods have been put forward to capture
useful information regarding gravitational lensing, with the lens being massive stellar objects, such as galaxy clusters and black holes. 
Lensing from rotating sources have also been well studied in the literature. While many of these have focused on rotating black hole
space-times, some studies have also been devoted to rotating naked singularity solutions of GR \cite{Naked}. 
The purpose of this paper is to study gravitational lensing in rotating solutions of Einstein's equations that can realistically model 
rotating galactic space-times. 

Recall that  there are three popular lensing schemes available in the literature, i.e strong, weak, and micro lensing. 
If the total bending of light is large, then we are in the realm of strong lensing, which will be our focus here. 
In this regard, one of the most extensively used strong lensing formalisms in the background of static, spherically symmetric space-times (SSS)
was given by Virbhadra and Ellis in \cite{6}, \cite{7}, which was later generalized to stationary, axi-symmetric space-times (SAS). 
The Virbhadra-Ellis formalism depends on two basic assumptions: the space-time is asymptotically flat and the source as well as observer are at 
infinite distances away from the lensing object. 
 
If either the source or the observer (or both) are at finite distances from the lensing object where the space-time is not flat, 
a light ray will bend immediately after leaving the source. Moreover, if the space-time is not asymptotically flat, there will be finite non-trivial contributions to the bending angle of light due to the space-time curvature at infinity, which is distinct from that due to the lens. 
Under these circumstances, the Virbhadra-Ellis formalism may not be applicable in a useful manner. 
In such situations, an alternative strong lensing formalism developed by Perlick \cite{1} (see also \cite{2}, \cite{3}) is more useful. 
Perlick's strong lensing formalism does away with the previous assumptions and describes lensing in a somewhat different approach. 
Here, the lens equation has explicit functional dependence on the source and observer positions, implying that  
the source or the observer need not to be at asymptotically infinite distances. 
Perlick's formalism was originally derived for SSS. Here, we will generalize this to the case of SAS, which is the focus of this paper. 

With a generic lensing formalism for SAS, we will focus on rotating galactic models. 
In particular, our aim is to study strong lensing in rotating solutions of GR that can be thought of as realistic models for galaxies at least
far from the galactic center. We will study two such models here. The first one is a rotating generalization of a galactic model obtained by 
Bharadwaj and Kar (BK) in \cite{Sayan}, who considered dark matter with pressure as a general relativistic model of galactic halos. 
The metric was obtained under the assumption of flat galactic rotation curves and was matched to an external Schwarzschild solution. 
Specifically, BK considered emissions from neutral Hydrogen (HI) clouds and used the observational fact that these lead to flat galactic 
rotation curves with circular velocity $v_c$, such that $v_c/c\sim 10^{-4}$ with $c$ being the speed of light. Using the fact that these
HI clouds move in the galactic plane under the gravitational forces of the halo region, an analytic form of $v_c$ was then computed 
from a metric ansatz, and the metric components were then calculated in terms of $v_c$, after matching with an external 
Schwarzschild solution. This was perhaps the first computation of
a galactic metric directly using observational data. The BK metric is not a vacuum solution, and the matter has non-zero radial and tangential
pressures. This was attributed to the dark matter content of galactic halos. 

Note that BK allows for a strictly constant circular velocity and obtains a static metric. A rotating metric obviously provides a more 
realistic galactic model, given the dark matter interpretation above. Starting directly with a metric ansatz involving rotation 
and then deriving a galactic metric ab-initio proved difficult, and we 
take a more conventional point of view, namely, taking the BK metric as a seed metric, we use the Newman-Janis Algorithm (NJA) \cite{9}. 
This is a standard technique in GR to obtain stationary space-times from their analogous static cousins. 
Here, we use the NJA to obtain a rotating BK (RBK) solution. In a realistic situation, one would like
to match this with a Kerr solution at a matching radius that signifies the galactic radius. Using standard junction conditions, we will show that it
is possible to match the RBK solution to a Kerr metric {\it without a slowly rotating approximation}. 
The RBK metric always satisfies the weak energy conditions away from the galactic
center, and therefore seems reasonable. One can of course ask what happens to the circular velocity in the RBK model. We compute this and
show that far from the galactic center, the rotation curves are reasonably flat, making the RBK metric an attractive rotating galactic model. 

We next move on to Bertrand space-times (BSTs), a solution of GR that has a naked singularity at the origin, but nonetheless might 
provide an interesting galactic model. BSTs were discovered by Perlick in \cite{5} and are defined to be solutions of Einstein's equations
that admit closed stable orbits at each spatial radius. This is a GR version of the well known Bertrand's theorem in classical 
mechanics \cite{goldstein}. The fact that
celestial objects at least far from the galactic center move in roughly closed circular orbits makes BSTs viable candidates for
galactic dark matter. In \cite{11}, it was shown that the weak energy conditions are satisfied by matter seeding BSTs, in \cite{12}, the nature
of the circular velocity of test particles in the background of BSTs were contrasted with experimental data, and in \cite{10}, strong gravitational 
lensing from BSTs were discussed in the Virbhadra-Ellis formalism. All these computations were done for the static BST. 

In this context, it is of theoretical importance to ask if there is a rotating version of BSTs (RBSTs), i.e if one can construct a stationary axi-symmetric
solution in GR where each spatial radius admits closed circular orbits. Here we show that this is indeed possible, by applying the NJA to
Perlick's static BST. We will also see that the weak energy conditions are satisfied away from the central naked singularity. 
RBSTs thus might be an example of a viable galactic metric. However, as a caveat we point out that 
it is not possible to match a Kerr metric to RBSTs, even in the slow rotation approximation, for a finite matching radius, 
and we have to thus deal with a space-time that is 
not asymptotically flat, but has a conical defect at asymptotic infinity. Thought of as a possible space-time for galactic dark matter, 
we thus have to deal with a situation in which we have diffused dark matter extending to a large spatial radius. 

Having established the rotating Bharadwaj-Kar metric and rotating Bertrand space-time as realistic galactic metrics, we study strong lensing
phenomena in these. We use the generalization of Perlick's construction of strong lensing discussed before. In the first case, we point out the
possible differences that might arise from purely Kerr lensing, and analyse their observational significance. In the second case,
we obtain results for gravitational lensing by rotating naked singularities, and this complements the analysis of \cite{Naked}, although we point
out that the latter was done using a different strong field limit developed by Bozza \cite{Bozza}. \footnote{In this context, we should point out that
a naked singularity itself can be further classified into two categories: weakly naked and strongly naked \cite{6}. A weakly naked singularity 
is covered by one or more photon spheres \cite{7},\cite{8}, wheres no photon sphere exists for a strongly naked singularity. 
In view of this classification, BSTs are strongly naked.}

This paper is organised as follows. In section 2, we present the details of the derivation of the rotating Bharadwaj-Kar metric
and the rotating Bertrand space-time metric using the Newman-Janis algorithm. For both these, we provide the necessary checks for the
weak energy condition. The issue of circular velocity is also discussed in some details. 
In section-3, we present the necessary generalization of strong lensing applied to the case of stationary axi-symmetric space-times
 based upon Perlick's original prescription of \cite{1}. Section 4 consists of the formulation and analysis of gravitational lensing 
 in the background of RBK and RBST. Finally, section 5 concludes the paper with a summary and discussion of the main results.
Possible caveats in our analysis are also pointed out here. 


\section{Constructing rotating galactic metrics}
\label{sec-2}

In this section, we will present two possible solutions of rotating galactic metrics. As pointed out in the introduction, these are important
and interesting in their own rights as they can be interpreted as space-times seeded by rotating dark matter. We will first show the construction
of the rotating generalization of the BK metric \cite{Sayan}, and then move on to generate rotating 
solutions of Bertrand space-times proposed by Perlick \cite{5}. 

\subsection{Rotating BK Space-times}
\label{sec-2.1}

In \cite{Sayan}, a possible galactic metric was put forward, based on the observational evidence of flat galactic rotation curves. 
The authors started with a metric ansatz of the form 
\begin{equation}
ds^2 = - e^{2\phi(r)}dt^2 + e^{2\lambda(r)}dr^2 + r^2d\Omega^2
\end{equation}
with $d\Omega^2$ being the standard metric on the unit two sphere with coordinates $\theta$ and $\phi$. 
Constancy of the circular velocity defined as $v_c = L/r$ with $L$ being the conserved angular momentum, determined the function 
$\phi(r)$, which matching with an external Schwarzschild solution fixed the functional form of $\lambda(r)$. with a particular choice
of the equation of state for matter. With $v_c/c\sim 10^{-4}$ so
that higher powers of this quantity could be neglected,\footnote{Specifically, in \cite{Sayan}, terms up to ${\mathcal O}(v^2/c^2)$ were
retained, and higher order terms were dropped. This will be our level of approximation here as well}  
the final form of the metric obtained in \cite{Sayan} (a small error in that
paper necessitated a few modifications that were pointed out in \cite{SayanError}) is given by (after setting $c=1$)
\begin{equation}
\phi(r) = v^2\left(\log(r/R)-1\right)~,~~
\lambda(r) = \frac{v^2}{w+1} \left(w \left(\frac{r}{R}\right)^{-\frac{w+1}{w}}+1\right)
\label{KBnonrot}
\end{equation}
Here, $R$ is the matching radius with an external Schwarzschild solution. 
Taking the above as the seed metric, we apply the Newman-Janis algorithm to generate a rotating solution. The algorithm
involves a number of mathematical steps, which was proposed in \cite{9}, and is by now standard. 
We skip the details here (which can for example be found in the excellent article of Drake and Turolla \cite{Drake}), and simply
mention that following the necessary steps of the NJA, we get the stationay axi-symmetric metric for the RBK solution as 
$ds^2 = g_{\mu\nu}dx^{\mu}dx^{\nu}$, with 
\begin{eqnarray}
g_{tt} &=& -e^{2\phi(r)}~,~~g_{rr} = \frac{\Sigma}{\Delta}~,~~g_{\theta\theta}=\Sigma~,~~g_{t\phi}=
a\sin^2\theta\left(e^{2\phi(r)} - e^{\lambda(r) + \phi(r)}\right)~,\nonumber\\
g_{\phi\phi} &=& \sin^2\theta \left(a^2 \sin^2\theta \left(2 e^{\lambda(r) +\phi(r) }-
e^{2\phi(r) }\right)+\Sigma \right)
\label{RBKMetric}
\end{eqnarray}
Here, $a$ is the rotation parameter (identified with the angular momentum per unit mass) 
with $\phi(r)$ and $\lambda(r)$ given from Eq.(\ref{KBnonrot}). We have also defined 
\begin{equation}
\Delta = e^{-2 \lambda(r)} \left(a^2 \cos^2\theta+r^2\right)+a^2 \sin^2\theta~,~~
\Sigma = r^2 + a^2\cos^2\theta
\end{equation}

Having established the rotating BK solution, an important question is whether this can be matched with a Kerr external solution on
a time-like hypersurface. The Kerr metric, in Boyer-Lindquist coordinates is given by 
\begin{eqnarray}
g_{tt} &=& -\left(1-\frac{2 M r}{a^2 \cos^2\theta+r^2}\right)~,~~g_{rr} = \frac{a^2 \cos^2\theta+r^2}{a^2-2 M r+r^2}~,
~~g_{\theta\theta}=a^2 \cos^2\theta+r^2~,\nonumber\\
g_{t\phi} &=& -\frac{2 a M r \sin^2\theta}{a^2 \cos^2\theta+r^2}~,~~
g_{\phi\phi}=\sin^2\theta \left(a^2+r^2+\frac{2 a^2 M r \sin^2\theta}{a^2 \cos^2\theta+r^2}\right)
\label{KerrMetric}
\end{eqnarray}

We choose the same coordinates for the RBK and the Kerr solutions, and in order to match the solutions at the boundary without a 
boundary stress tensor, we need to check if the first and second fundamental forms can be matched on two sides of a time-like 
hypersurface $r=R$. As mentioned in \cite{Drake}, this hypersurface is an oblate-spheroid and is natural in axially symmetric solutions
of the Einstein equations. We write the Kerr metric in terms of the dimensionless variable $M_0=M/R$ with $R$ being the matching
radius (this is the same $R$ that appears in Eq.(\ref{KBnonrot})).\footnote{We are implicitly assuming that $R$ is in some sense
the halo radius of the galaxy. This should be treated as a simplifying assumption} 
We find that the induced metrics obtained from eqs.(\ref{RBKMetric})
and (\ref{KerrMetric}) indeed match with the choice 
\begin{equation}
M_0=\frac{\left(1-e^{-\frac{2 v^2}{c^2}}\right) \left(a^2 \cos^2\theta+R^2\right)}{2 R^2}
\label{M0sol}
\end{equation}
Note that matching of the two metrics imply that the rotation parameter $a$ is the same for both the RBK and Kerr space-times.
Using this value of $M_0$, we can compute the extrinsic curvature $K_{ab}$ $(a,b = t,\theta,\phi)$ on a time-like hypersurface
at $r=R$ due to the RBK and Kerr metrics. Across this hypersurface, up to order $v^2$, we find that (with $x=\cos\theta$),
\begin{eqnarray}
[K_{tt}] &=& -\frac{2a^2v^2x^2\left(a^2+R^2\right)^{1/2}}{R\left(R^2+a^2x^2\right)^{3/2}}~,~~
[K_{t\phi}]=\frac{2a^3v^2x^2\left(1-x^2\right)\left(a^2+R^2\right)^{1/2}}{R\left(R^2+a^2x^2\right)^{3/2}}~,~\nonumber\\
\left[K_{\phi\phi}\right] &=& -\frac{2a^4v^2x^2\left(1-x^2\right)^2\left(a^2+R^2\right)^{1/2}}{R\left(R^2+a^2x^2\right)^{3/2}}
\label{matchingEC}
\end{eqnarray}
Where $[\cdots]$ is the difference in the extrinsic curvatures on the two sides. We note first that on the equatorial plane $x=0$,
the matching of the extrinsic curvatures is valid for all values of $a$, to the order of approximation in $v/c$ that we consider. 
We now argue that this is also the case away from the equatorial plane. To see this, we note that the matching of the RBK and
the Kerr metrics necessarily require the identification of the rotation parameter of both. Then the fact that $M=M_0 R=a/k$ with $0<k<1$
as follows from the Kerr metric translates into an equation involving the r.h.s of Eq.(\ref{M0sol}). The solution yields
\begin{equation}
a = kR\left(\frac{v^2}{c^2}\right) + {\mathcal O}\left(\frac{v^4}{c^4}\right)
\label{aval}
\end{equation}
Inserting Eq.(\ref{aval}) into Eq.(\ref{matchingEC}) then yields exact matching (for any value of $a$) of the RBK and Kerr metrics
up to the desired order for generic values of the azimuthal angle. 

Next, we turn to the energy conditions in the RBK metric. The computations here are again standard -- as is common in rotating solutions 
of GR, we choose a locally flat tetrad basis and apply a transformation to render the stress tensor diagonal. At this point, we make specific
choices of the azimuthal angle $\theta$ and the equation of state parameter $w$ to simplify the analysis. We restrict ourselves to the equatorial plane and
choose $\theta = \pi/2$. We also choose $w=-1/2$ (note that $-1<w<0$ \cite{SayanError}). With this choice, we find that the weak
energy conditions are indeed satisfied away from $r=0$, i.e in the region of validity of the RBK metric. The expressions for the
density and the principal pressures are complicated for generic radii, but simplify near the matching surface $r=R$, which we show
for completeness :
\begin{eqnarray}
\rho &=& \frac{4 v^2 (R-r)}{R r^2}+\frac{105 a^2 v^2 (R-r)}{4 R^5}+\frac{9 a^2 v^2}{4 R^4}~,~
P_r=-\frac{v^2 \left(9 a^2+2 R^2\right) (R-r)}{R^5}-\frac{2 a^2 v^2}{R^4}\nonumber\\
P_{\theta} &=& \frac{v^2 \left(R^2-a^2\right)}{R^4}+\frac{v^2 \left(R^2-5 a^2\right) (R-r)}{R^5}~,~
P_{\phi} = \frac{v^2 \left(129 a^2+4 R^2\right) (R-r)}{4 R^5}+\frac{v^2}{R^2}+\frac{13 a^2 v^2}{4 R^4}
\end{eqnarray}
These of course reduce to the expressions for the components of the static BK metric in the limit $a\to 0$. Note that
in this case, the radial pressure is negative. As explained in \cite{Sayan},\cite{SayanError}, this is not uncommon when 
one considers the fluid matter to be sourced by a real scalar field, when the field is at a minimum value. 

Finally, we will make a few remarks about the circular velocity of massive particles (that do not back-react on the metric) in RBK space-times. 
This is not difficult and we will simply state the final result that following the steps of \cite{BardeenKerr}, we obtain, near the matching
radius the circular velocity $v_c = L/r$ where $L$ is the conserved angular momentum along the orbit,
\begin{equation}
v_c \sim v + \frac{a^2v}{r} + \frac{\left(R-r\right)\left(5a^3 + 9aR^2\right)v^2}{2R^4} + \frac{a\left(a^2+3 R^2\right)v^2}{R^3} 
\end{equation}
The constantcy of $v_c$ can be then gleaned from Eq.(\ref{aval}) up to ${\mathcal O}\left({v^2/c^2}\right)$. 

\subsection{Rotating Bertrand Space-times}

As mentioned in the introduction, Bertrand space-times (BSTs) are solutions of Einstein equations that admit closed stable orbits at
each radial distance, and were discovered by Perlick in \cite{5}. We consider, in this paper, 
a specific class of BSTs given by the metric 
\begin{equation}
ds^2 = -\frac{dt^2}{G + \alpha \sqrt{r^{-2} + K} } + \frac{dr^2}{ \beta ^2 \left( 1 + K r^2 \right) } 
+ r^2 \left( d \theta ^2 + \sin ^2 \theta d \phi ^2 \right) \label{a8}
\end{equation}
where $ G $, $ K $ and $ \alpha $ are positive real constants and $ \beta $ is a positive rational number. 
Here, $ \alpha $ has the dimension of length, $ K $ has dimension of length squared inverse and $ G, \beta $ are 
dimensionless quantities. For simplicity, we choose $ K=0 $, which transforms metric (\ref{a8}) in this simpler form.
\begin{equation}
ds^2 = -\frac{dt^2}{G + \frac{\alpha}{r} } + \frac{dr^2}{ \beta ^2 } + r^2 \left( d \theta ^2 + \sin ^2 \theta d \phi ^2 \right) \label{a9}
\end{equation}
The above metric can also be re-written in the following equivalent form
\begin{equation}
ds^2 = -\frac{dt^2}{1 + \frac{ r_h }{r}} + \frac{dr^2}{ \beta ^2 } + r^2 \left( d \theta ^2 + \sin ^2 \theta d \phi ^2 \right) \label{a10}
\end{equation}
where $ r_h= \frac{\alpha}{G} $ and the time coordinate has been rescaled as, $ t/\sqrt{G} \rightarrow t $. If we think of BSTs as
galactic metrics at least far from the galactic center \cite{12}, $ r_h $ can be 
related to the galactic length scale, and $ \beta $ fixes the nature of closed orbits, with $ \beta = 1 $ 
representing Keplerian orbits \cite{5}. Moreover, the constant time hyper-surfaces at special infinity have conical defects 
which are specified by $ \beta $, with $ \beta=1 $ corresponding to zero conical defect or flat space.  
Again, it can be shown that we require $ 0< \beta <1 $ to satisfy 
WEC, \cite{10}, \cite{11}. From now on, by Bertrand space-time (BST), we shall always correspond to the space-time with metric (\ref{a10}).

We shall take BST as a seed metric and apply Newman-Janis Algorithm (NJA) to it to obtain its rotating counterpart. 
Note that BSTs were obtained by Perlick by considering a generic static spherically symmetric solution of Einstein's equations
and imposing conditions on the stability of circular orbits. This route is difficult to envisage for stationary solutions of GR. This is
the reason we resort to the NJA. 
Then, a straightforward computation yields the rotating generalisation of Eq.(\ref{a10}) as 
\begin{equation}
	ds^2 = - A(r,\theta) dt^2 + B(r,\theta) dr^2 + C(r,\theta) d\phi^2 + F(r,\theta) d\theta^2 - 2D(r,\theta) dt d\phi \label{a7}
\end{equation}

\[ \text{where} ~~~ A(r,\theta) = \frac{1}{1+ (r_h r)/\rho^2}, ~~~ B(r,\theta) = \frac{\rho^2}{\Delta}, ~~~ F(r,\theta) = \rho^2, \]

\[ D(r,\theta) = \left(\frac{1}{\beta \sqrt{1 + (r_h r)/\rho^2}}-\frac{1}{1+ (r_h r)/\rho^2}\right)a \sin^2\theta, \]

\[ \text{and} ~~~ C(r,\theta) = \left[(r^2+a^2)+\left(\frac{2}{\beta \sqrt{1+ (r_h r)/\rho^2}} - 1 - \frac{1}{1+ (r_h r)/\rho^2}\right)a^2 \sin^2\theta \right]\sin^2\theta \]

\[ \text{with} ~~~ \rho^2 = r^2 + a^2 \cos^2 \theta ~~~ \text{and} ~~~ \Delta = \rho^2 \beta^2 + a^2 \sin^2 \theta. \]
Note that constant time hyper-surfaces at spatial infinity have conical defects for RBSTs as is the case with BSTs. 
Let's now consider the nature of singularity of RBST. By calculating different invariant scalars it can be shown that RBST 
has an essential singularity at $ r=0 $. The position of event horizon for the type of rotating metric (\ref{a7}) is determined from the 
condition, $ B(r,\theta) = \infty $, which in the case of RBST, transforms to the condition, $ \Delta = 0 $. 
Since $ \Delta (=\rho^2 \beta^2 + a^2 \sin^2 \theta) $ is found to be the sum of only positive quantities for RBST, it can never be zero. 
Therefore, RBST does not contain any event horizon and the singularity at $ r=0 $ is naked. 
It is important to remember that normal BST has a strongly naked singularity at $ r=0 $ and there is no photon sphere for BST. 

To determine if it is a weakly or strongly naked singularity, 
we need to analyze the metric in detail. We shall consider the equatorial plane $ (\theta=\frac{\pi}{2}) $ only in this paper.
On this plane, the form of the metric (\ref{a7}) transforms to

\begin{equation}
ds^2 = - A(r) dt^2 + B(r) dr^2 + C(r) d\phi^2 - 2D(r) dt d\phi \label{a13}
\end{equation}
where the metric coefficients take the following form

\[ A(r) = \frac{1}{1 + \frac{r_h}{r}}, ~~ B(r) = \frac{r^2}{r^2 \beta^2 + a^2}, ~~ D(r) = \left( \frac{1}{\beta \sqrt{1 + \frac{r_h}{r}}} - \frac{1}{1 + \frac{r_h}{r}} \right) a ~ , \]
\[ \text{and} ~~~ C(r) = r^2 + a^2 + \left( \frac{2}{\beta \sqrt{1 + \frac{r_h}{r}}} - 1 - \frac{1}{1 + \frac{r_h}{r}} \right) a^2 ~ . \]
The equation of a photon circle $ (r_{pc}) $ on the equatorial plane of a general rotating metric of the form (\ref{a13}) is given by

\begin{equation}
		\left. \Big[ A(r) \Big( A(r) C'(r) - A'(r) C(r) \Big) + 2 \Big( A(r) D'(r) - A'(r) D(r) \Big) \left( D(r) - \sqrt{D^2 (r) + A(r) C(r)} \right) \Big]\right \vert_{r= r_{pc}} = 0 \label{a14}
\end{equation}

If we put $ a=0 $ or $ D(r)=0 $ in Eq.(\ref{a14}), we shall get back the standard photon sphere equation for static space-times,
i.e $A(r)/A'(r)=C(r)/C'(r)$.
Putting the above mentioned expressions of the metric coefficients in Eq.(\ref{a14}), it can be shown that RBSTs do have photon 
circles for light rays co-rotating with the rotation of the space-time. There are two types of motion of photons in rotating space-times. 
Photons which are co-rotating with the rotation of space-time (direct) and photons which are counter-rotating (retrograde). 
Due to its rotation, the space-time will carry the light rays along with itself so that they will be forced to bend in the direction of rotation. 
In absence of rotation, they would have not bent that way. As a result, for direct rays, the bending will be more as compared to the 
corresponding rays in static background and for retrograde rays, the bending will be less. Therefore, it seems physically 
very difficult to have a rotating space-time without photon circles for direct rays.

On the other hand, we cannot in general apply this statement to retrograde photons for rotating space-time. 
In case of RBST, we find that there is indeed no photon circles for retrograde rays. Eq.(\ref{a14}) does not give any real solution 
for $ r $, for counter-rotating light rays. This establishes the fact that RBST retains strongly naked singular character for retrograde light rays. 
To summarize, the RBST is an example of rotating naked singularity which is weakly naked for direct (co-rotating) light rays and 
strongly naked for retrograde (counter-rotating) rays. A more detailed analysis of lensing for both direct and retrograde rays is 
performed in section 4 below.

Now, recall that the existence of stable circular orbits at each radial distance was the defining
property of BSTs. So the natural question arises is whether there exists
stable circular orbits at each value of the radius in RBST. For this, we use the
normalization condition of the 4-velocity for timelike geodesics to get 
$\dot{r}^2 + V_{eff}(r)=0 $ where
\begin{equation}
V_{eff}(r) = \frac{1}{B(r)}[- A(r) \dot{t}^2 + C(r) \dot{\phi}^2 - 2D(r)
\dot{t} \dot{\phi}+1]
\label{t}
\end{equation}
where an over-dot represents differentiation with respect to the affine
parameter along the time-like geodesic. Here $\dot{t}$ and $\dot{\phi}$
can be written in terms of the two constants of motion $L$ and $E$ as
\begin{equation}
        \dot{t} = \frac{ C(r) E - D(r) L }{ D(r)^2 +A(r) C(r) }, ~~~~~~~~
\dot{\phi} = \frac{D(r) E + A(r) L }{ D(r)^2 + A(r) C(r) } \label{new}
\end{equation}
Now, as is standard, we introduce the inverse distance $ u = \frac{1}{r} $ and the variable $x=L-a E$
where $a$ is the rotation parameter and solve $V_{eff}(u)=0$,
$V_{eff}'(u)=0$ to find  $E$ and $x$ for fixed values of $u$, $a$, $\beta$
and $r_h$ (the prime denotes differentiation with respect to $u$). Those
solutions (with $E>0$ and $x$ having both positive and negative values) for
which
$V_{eff}''(u)>0$ is satisfied, gives a stable circular orbit at that value
of $u$. Thus varying $u$ and continuing this process we can find out the
region of validity for the stable circular orbits in RBST. Since analytical computations become cumbersome,
the results of this analysis is best presented graphically. 
\begin{figure}[t!]
        \centering
        \begin{subfigure}{.45\textwidth}
                \includegraphics[width=\textwidth]{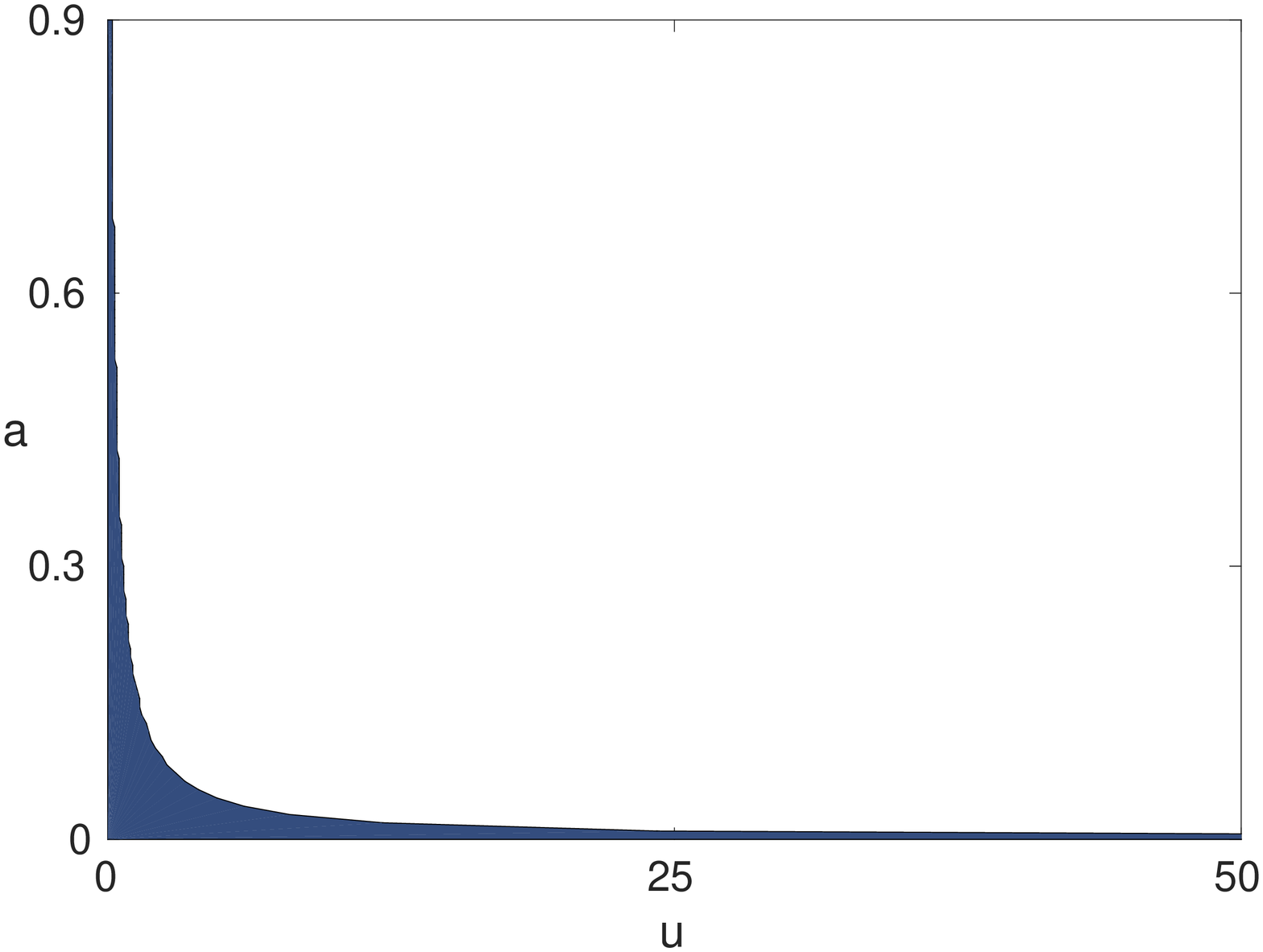}
                        \caption{ $ \beta=0.1 $}
        \end{subfigure}\hspace{0.1cm}
        \begin{subfigure}{.45\textwidth}
                \includegraphics[width=\textwidth]{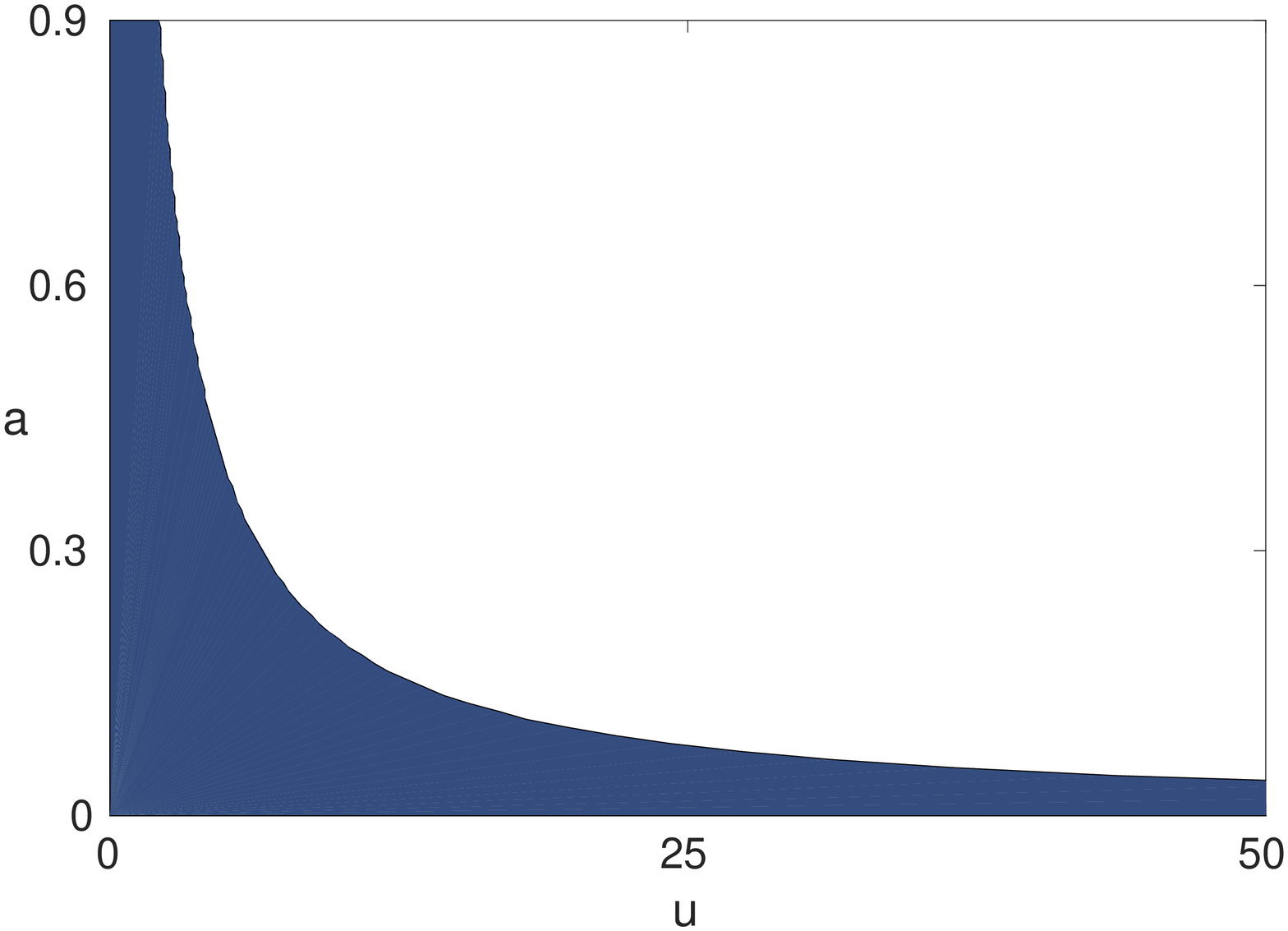}
                        \caption{ $ \beta=0.9 $}
        \end{subfigure}
        \caption{Plots of rotation parameter `$ a $' vs inverse distance `$ u $'. }
        \label{fig1}
\end{figure}

Figure (\ref{fig1}) shows region plots between the rotation
parameter, $ a $ and the inverse distance, $ u $ for two values of $ \beta=0.1$ and $0.9 $,
where the colored regions represent the points in the parameter space for which there
exists stable circular orbits. The
value of $ r_h = 100 $ is taken to be fixed for both the plots. The conclusions from this 
analysis can be summarized as follows. For a fixed value of $\beta$ and given value of $u$ (or $r$), we find that 
stable circular orbits exist only below a certain value of the rotation parameter $a$.\footnote{In the static case
$a=0$, stable circular orbits exist for all values of the radial distance for all $\beta$.} This value of $a$ increases
as $u\to 0$ or $r \to \infty$. The maximum value of $a$ for a given radial distance can be enhanced by increasing the 
value of $\beta$. Thus we conclude that in principle there can be stable circular orbits at all values of the radial distance
with a given maximum value of $a$ that depends on the parameter $\beta$. In this sense, RBSTs of Eq.(\ref{a13}) 
generalize BSTs to an axially-symmetric scenario.

\section{Lens equation for Stationary, Axi-symmetric Space-times}
\label{sec-3}

Our next aim would be to study strong gravitational lensing in the backgrounds of the RBK and RBST metrics discussed in
the previous section. In particular, we will use the formalism of Perlick \cite{1}, after appropriately generalizing it to a 
rotating background. We will focus on the equatorial plane, as the computations become extremely tedious for generic values
of the azimuthal angle. 

\begin{figure}[h]
\begin{center}
\includegraphics[width=0.6\textwidth]{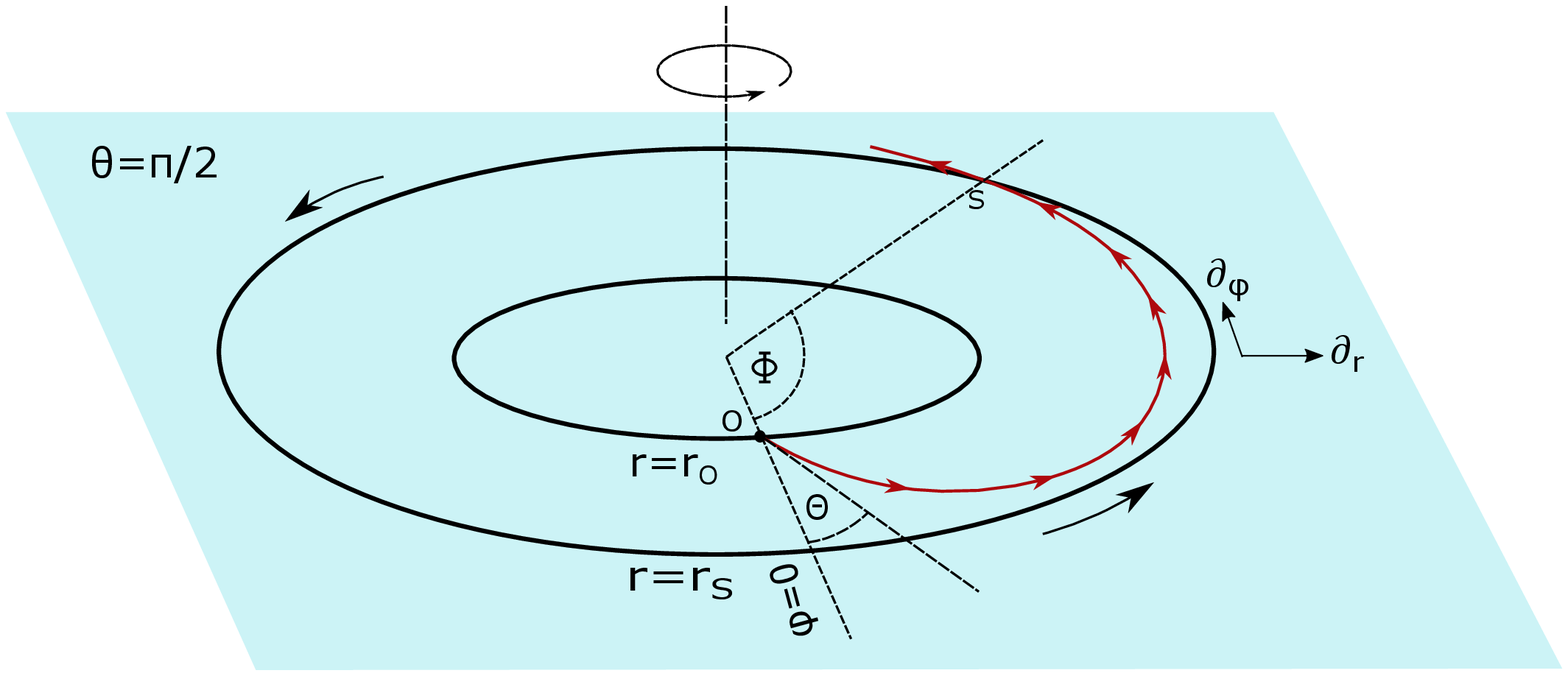}
\caption{In this figure, the observer is situated at $ r=r_O $, $ \phi=0 $ in the equatorial plane i.e. $ \theta=\pi/2 $. The light sources are 
distributed over the circle $ r=r_S $. The location of a particular source point on the source circle is given by its angular position $ \Phi $. 
The observer will see image of the same source point at angular position $ \Theta $ on the observer's sky, as shown. This figure is 
inspired from FIG.1 of \cite{1}.}
\label{fig2}
\end{center}
\end{figure}

The form of the most general stationary, axially-symmetric metric is given in Eq.(\ref{a7}) and the corresponding reduced form of 
the metric on the equatorial plane is given in Eq.(\ref{a13}). Let us consider, in Fig.(\ref{fig2}), a ray diagram of light bending to study 
the characteristics of strong lensing phenomena in the background of stationary axi-symmetric space-times (SAS).

The lens equation analytically corresponds to an equation of the form $ \mathcal{F}(\Theta , \Phi)=0 $. The values of $ \Theta $ are 
restricted from $ -\pi $ to $ \pi $, wheres $ \Phi $ can take any value modulo $ 2\pi $. In principle, light rays will originate from the 
source circle ($ r=r_S $) at position $ \Phi $ at time $ t=0 $ (say), will follow null geodesics and reach the observer at $ r=r_O $, 
$ \phi=0 $ at a later time $ t=T>0 $ (say). But to derive the lens equation, it is convenient to choose the opposite motion. 
Let us consider the past oriented light rays originating from the observer at $ t=0 $ and move backward in time to reach at some 
point of the source circle at an earlier time $ t=-T<0 $. The tangents to these light rays at $ r=r_O $ make angles $ \Theta $ with 
the direction of $ \partial_r $, which represent the image positions in the observer's sky. The lens equation, therefore, gives the 
source position $ (\Phi) $ as a function of the image position $ (\Theta) $. Two important scenarios in this regard can take place.

\begin{itemize}
	\item For a given value of $ \Theta $, the existence of $ \Phi $ is not always guaranteed. This happens 
	if the light originating from the observer's location fails to reach the source circle itself, i.e., the light ray bends so 
	much that it curls into the center at $ r=0 $ without approaching towards the source circle.
	
	\item On the other hand, the uniqueness of $ \Phi $ for a given value of $ \Theta $ is also not always confirmed. 
	The uniqueness fails if the light ray meets the source circle at $ r=r_S $ more than once at different angles. In this situation, 
	all these different source points will be coincident on the same image point on the observer's sky. In other words, there will be 
	multiple images at the same point -- one behind the other. These images which are covered by others are known as 
	``hidden images". 
\end{itemize}
The light-like geodesics in the background of the metric in Eq.(\ref{a13}) will be solutions of Euler-Lagrange equations derived from 
the following Lagrangian :
\begin{equation}
	\mathcal{L} = \frac{1}{2} \left[ - A(r) ~ \dot{t}^2 + B(r) ~ \dot{r}^2 + C(r) ~ \dot{\phi}^2 - 2D(r) ~ \dot{t} ~ \dot{\phi} \right] \label{r}
\end{equation}
where an over-dot represents differentiation with respect to the affine parameter along the light-like geodesic ($ s $). It
allows us to define two constants of motion corresponding to the time translation and azimuthal symmetries. They are given, respectively, as
\begin{equation}
	- A(r) ~ \dot{t} - D(r) ~ \dot{\phi} = -E, ~~~~~~~~~~~ - D(r) ~ \dot{t} + C(r) ~ \dot{\phi} = L \label{s}
\end{equation}
where $ E $ and $ L $ represent energy and angular momentum of photon respectively. Expressing $ \dot{t} $ and 
$ \dot{\phi} $ in terms of $ E $ and $ L $, we obtain Eq.(\ref{new}). 
Since the rest mass of photon is zero, the Lagrangian of Eq.(\ref{r}) for light-like geodesics becomes
\begin{equation}
	 - A(r) ~ \dot{t}^2 + B(r) ~ \dot{r}^2 + C(r) ~ \dot{\phi}^2 - 2D(r) ~ \dot{t} ~ \dot{\phi} = 0 \label{u}
\end{equation}
Substituting $ \dot{t} $ in the above equation from the first expression of Eq.(\ref{s}), we get
\begin{equation}
	A(r) B(r) ~ \dot{r}^2 + \left( D(r)^2 + A(r) C(r) \right) ~ \dot{\phi}^2 - E^2 = 0 \label{v}
\end{equation}
Let us now choose the following initial conditions that will satisfy Eq. (\ref{v}) at the observer's position (initial position) 
from where the light ray originates.
\begin{equation}
	r \vert_{s=0} = r_O, ~~~~ \dot{r} \vert_{s=0} = -\frac{E  \cos \Theta}{ \sqrt{A(r_O) B(r_O)} } ~ ; \label{w}
\end{equation}
\begin{equation}
	\phi \vert_{s=0} = 0, ~~~~ \dot{\phi} \vert_{s=0} = -\frac{E  \sin \Theta}{ \sqrt{D(r_O)^2 + A(r_O) C(r_O)} } \label{x}
\end{equation}
Using Eq.(\ref{x}) and the first expression of Eq.(\ref{s}), we find
\begin{equation}
	t \vert_{s=0} = 0, ~~~~ \dot{t} \vert_{s=0} = \frac{ D(r_O) E  \sin \Theta}{ A(r_O) \sqrt{D(r_O)^2 + A(r_O) C(r_O)} } + \frac{ E }{A(r_O)} \label{y}
\end{equation}
We can determine $ L $ from the second expression of Eq.(\ref{s}) using the initial conditions of 
$ \dot{t} $ and $ \dot{\phi} $ as
\begin{equation}
	 L = - D(r_O) ~ \dot{t} \vert_{s=0} + C(r_O) ~ \dot{\phi} \vert_{s=0} 
	\Rightarrow ~~ L = -\frac{ \sqrt{D(r_O)^2 + A(r_O) C(r_O)} }{ A(r_O) } ~ E \sin \Theta - \frac{ D(r_O) }{ A(r_O) } ~ E \label{a5}
\end{equation}
Now putting the expression of $ \dot{\phi} $ from Eq. (\ref{new}) into Eq.(\ref{v}), yields
\begin{equation}
	A(r) B(r) ~ \dot{r}^2 = E^2 - \frac{ \left( D(r) E + A(r) L \right)^2 }{ D(r)^2 + A(r) C(r) } \label{z}
\end{equation}
If $ \dot{r} $ does not change sign
\begin{equation}
	\dot{r} = \frac{dr}{ds} = \frac{1}{ \sqrt{A(r) B(r)} } \sqrt{ E^2 - \frac{ \left( D(r) E + A(r) L \right)^2 }{ D(r)^2 + A(r) C(r) } } \label{a1}
\end{equation}
Again, from the second expression of Eq.(\ref{new}), we have
\begin{equation}
	\dot{\phi} = \frac{d\phi}{ds} = \frac{D(r) E + A(r) L }{ D(r)^2 + A(r) C(r) } \\ \label{a2}
\end{equation}
Combining Eq.(\ref{a1}) and Eq.(\ref{a2}), we obtain
\begin{equation}
    \frac{d\phi}{dr} = \frac{ \sqrt{A(r) B(r)} \left( D(r) E + A(r) L \right) }{ \sqrt{D(r)^2 + A(r)C(r)} \sqrt{E^2 \left( D(r)^2 + A(r)C(r) \right) - \left(  D(r) E + A(r) L \right)^2} } \\ \label{a3}
\end{equation}
From the above equation we get the desired Lens equation in integral form
\begin{equation}
\Phi(\Theta) = \int_{r_O}^{r_S} \frac{ \sqrt{A(r) B(r)} \left( D(r) E + A(r) L(\Theta) \right) ~ dr }
{ \sqrt{D(r)^2 + A(r)C(r)} \sqrt{E^2 \left( D(r)^2 + A(r)C(r) \right) - \left(  D(r) E + A(r) L(\Theta) \right)^2} } \label{a4}
\end{equation}

Here, the $ \Theta $ dependence of $ \Phi $ enters through $ L $ which is a function of the source position in the observer's 
sky ($ \Theta $). Similarly, if $ \dot{r} $ changes sign along the geodesic from $ r_O $ to $ r_S $, the integral in Eq.(\ref{a4}) 
has to be replaced by a piece-wise integration. As far as the constant value of $ E $ is concerned, we can always make $ E=1 $ 
with a proper suitable choice of the affine parameter (see Eq.(8) of \cite{4}). Since we are considering a past oriented lightlike geodesic, we must use a negative sign before $ E $. Therefore, the above equation, with $ E=-1 $, reduces to
\begin{equation}
	\Phi(\Theta) = \int_{r_O}^{r_S} \frac{ \sqrt{A(r) B(r)} \left( A(r) L(\Theta) - D(r) \right) dr}{ \sqrt{D(r)^2 + A(r)C(r)} \sqrt{D(r)^2 + A(r)C(r) - \left( A(r) L(\Theta) - D(r) \right)^2} } \label{a6}
\end{equation}

This is the master equation that we use in the rest of our analysis, and generalizes the static situation of Perlick given in \cite{1}.
We shall use this equation to calculate $ \Phi $ as a function of $ \Theta $ and study the lensing phenomena for SAS. 
The importance of the above equation lies in the fact that it is exact without any asymptotic assumptions. 
Hence, it can be used to analyze gravitational lensing in the equatorial plane for any stationary space-time having metric 
of the form given in Eq.(\ref{a13}). If we set $ D(r)=0 $ (i.e. $ a=0 $) in Eq.(\ref{a6}), we obtain the corresponding 
lens equation for a static space-time, Eq.(14) of \cite{1}.


\section{Strong Lensing in Rotating Galactic Space-times}
\label{sec-4}

In this section, we present our results on strong lensing in rotating galactic space-times exemplified by the metrics that we have
constructed in section 2. 
\subsection{Strong Lensing in RBK Space-times}

In this subsection, we study strong lensing in RBK space-times using the formalism developed in the previous section. 
To make the analysis realistic, we take actual data from Sombrero Galaxy (NGC 4594) as an example for values of different 
parameters involving in the lensing calculation. We consider

\[ w=-\frac{1}{2}, ~~ v=2 \times 10^5 ~ \text{m s}^{-1}, ~~ R=7.51 ~ \text{Kpc}, ~r_O=9.53~ \text{Mpc}, ~~ r_S= \frac{4R}{5}, \]
All these data are collected from standard literature. Here, $ R $ is the radius of the outer edge of the galaxy where it 
matches with an external Kerr metric and $r_O$ is the distance from the earth to the Sombrero galaxy. 
After matching the first fundamental form between the two space-times at $ r=R $, 
the mass of the galaxy for an external observer is found to be, $ M= \frac{c^2 R}{2G} \left( 1-e^{-\frac{2v^2}{c^2}} \right) $ (this can 
be seen from Eq.(\ref{M0sol}) after restoring dimensions).
 
From the above values, we see that $ r_S < R < r_O $ in this case. Therefore, according to Fig.(\ref{fig2}) 
(interchanging the positions of $ r_S $ and $ r_O $), after originating from the observer circle, light rays with 
$ |\Theta|>\frac{\pi}{2} $ will first travel in the Kerr space-time, cross the matching radius and enter into the 
RBK space-time (which is our model for the Sombrero Galaxy), then traverse in it before reaching the source circle. 
So a portion of its path resides within the galaxy which provides us the characteristic signatures of lensing for the galaxy itself. 
And this non-trivial effect coming from the galaxy part will make the source position ($ \Phi $) different from what it would have 
been if the galaxy is modeled by Kerr space-time without matter. The longer the path length within the galaxy, the difference 
will be more prominent. We will explicitly calculate this difference for the value of the rotation parameter, 
$ a=0.1 \frac{MG}{c^2} $ (in units of length).

\begin{figure}[h]
	\begin{center}
		\includegraphics[width=0.5\textwidth]{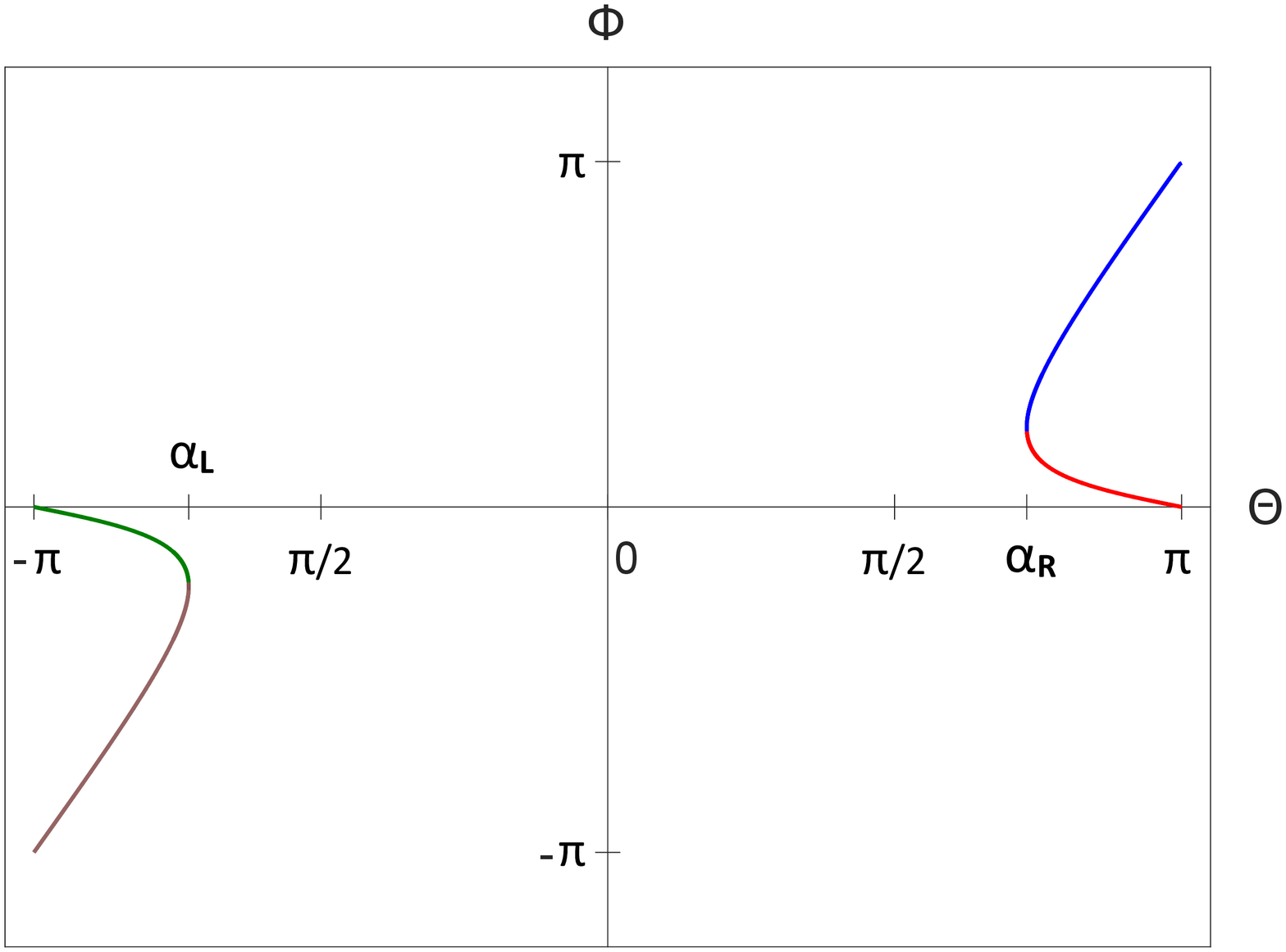}
		\caption{Representative lensing plot in RBK-Kerr space-time in the equatorial $ (\theta=\frac{\pi}{2}) $ plane 
		with matching radius $ R=10 $, $ r_O=12 $ and $ r_S=9 $. Here, $ \alpha_R $ (on the right) and $ \alpha_L $ (on the left) 
		are the minimum 
		initial angles for direct rays and retrograde rays respectively. Natural units with $ c=G=1 $ are adopted here. }
		\label{representative}
	\end{center}
\end{figure}

Before going into details of the exact analysis, we first discuss some characteristic features of 
the $ \Phi-\Theta $ lensing plots for the $ r_S < r_O $ case under consideration. Since the observer circle is 
located outside the source circle, light rays with $ |\Theta|< \frac{\pi}{2} $ will move towards spatial infinity 
without ever reaching the source circle. So there will be no image for these rays. If the initial angle 
$ |\Theta| $($ >\frac{\pi}{2} $) is increased continuously, there will be a specific angle for which the 
corresponding light ray will just reach the source circle tangentially. Since the rotation of the space-time breaks the 
spherical symmetry in stationary space-times, this angle will be different for direct 
(co-rotating with $ \frac{\pi}{2} < \Theta < \pi $) rays and retrograde 
(counter-rotating with $ -\pi > \Theta > -\frac{\pi}{2} $) rays. In Fig.(\ref{fig2}), a direct light ray in red color is 
shown. Fig.(\ref{representative}) shows a representative lensing plot in RBK-Kerr space-time for 
$ r_S<r_O $ case \footnote{For the corresponding lensing plot in static, spherically-symmetric space-time with 
$ r_S<r_O $, the reader is referred to Fig.(4) of \cite{1}.}.

Fig.(\ref{representative}) is obtained numerically in the equatorial plane as the metric components of RBK are 
too complicated to integrate Eq.(\ref{a6}) analytically. Moreover, we cannot use Eq.(\ref{a6}) directly to obtain 
$ \Phi-\Theta $, as the $ r_S<r_O $ case has turning points ($ \dot{r}=0 $ positions) along the path of light rays. 
So we modified the lens equation into a piece-wise integral equation and then used this for numerical computation. 
As can be seen from Fig.(\ref{representative}), rays with initial angle greater than the minimum 
values ($ \alpha_R $ and $ \alpha_L $) will meet the source circle twice, producing two images at the same point 
in the observer's sky. Therefore, it will produce hidden images for each $ \Theta $. In the figure, lines with red and green 
colors correspond to real images, and lines with blue and brown colors correspond to hidden images, 
which are located behind the real images. Existence of hidden images are a generic feature for the $ r_S<r_O $ case. 
From now on, we shall use the following terminology and color scheme for describing different portions of Fig.(\ref{representative}).

\begin{itemize}
	\item \textbf{Right-branch, lower-half:} the portion in red. This represents real images for direct rays.
    \item \textbf{Right-branch, upper-half:} the portion in blue. This represents hidden images for direct rays.
    \item \textbf{Left-branch, lower-half:} the portion in green. This represents real images for retrograde rays.
    \item \textbf{Left-branch, upper-half:} the portion in brown. This represents hidden images for retrograde rays.
\end{itemize}
So in both the branches `lower-half' corresponds to real images and `upper-half' corresponds to hidden images.

\begin{figure}[t]
	\centering
	\begin{subfigure}{.45\textwidth}
		\includegraphics[width=\textwidth]{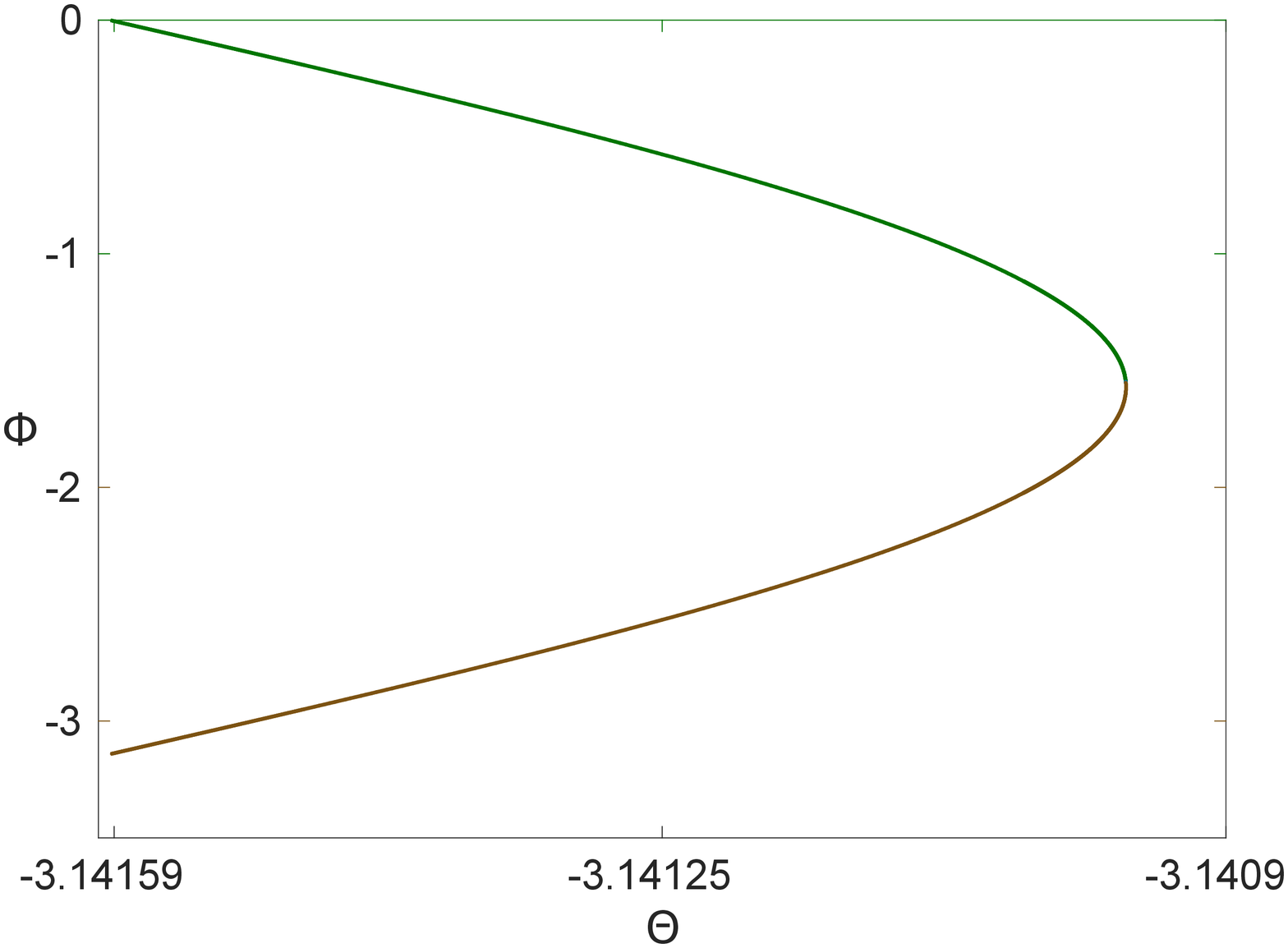}
		\caption{ Left-branch }
	\end{subfigure}
	\begin{subfigure}{.45\textwidth}
		\includegraphics[width=\textwidth]{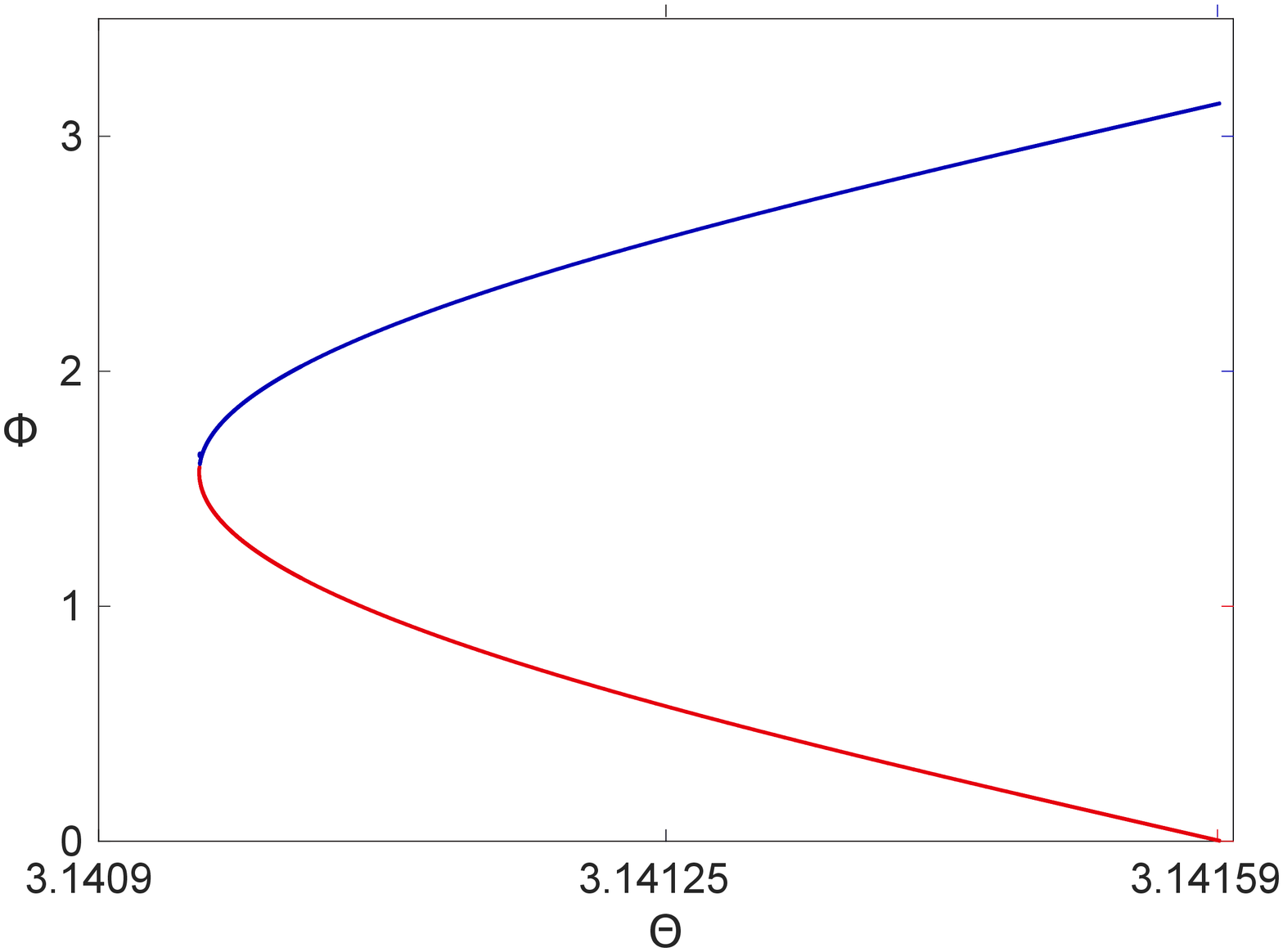}
		\caption{ Right-branch }
	\end{subfigure}
	\caption{Lensing plots for RBK-Kerr space-times with actual data taken from the Sombrero Galaxy. $ (a) $ shows the left
	 hand portion of the $ \Phi-\Theta $ plot and $ (b) $ shows the right hand portion. Here also, red and green parts are 
	 considered as `lower-half', and blue and brown parts are considered as `upper-half'. }
	\label{plotRBK}
	\end{figure}

Let us now discuss the lensing in RBK space-time considering the actual data as given in the beginning of this 
subsection. We assume that the Sombrero Galaxy represents our RBK space-time which is matched with an external
 vacuum space-time represented by the Kerr metric. Therefore, light travels in both the space-times during its 
 whole path and produces real as well as hidden images.\footnote{In computing the hidden images we are tacitly assuming
 that the RBK metric is valid all the way up to the photon circle. This assumption might be violated in a realistic scenario. However,
we will ignore this fact here as this calls for a more sophisticated analysis than the one reported in \cite{Sayan}. Note that no
such problem exists for the direct images.} 
As the radius of source circle is very small with respect to the 
 radius of observer circle ($ \frac{r_S}{r_O} \approx 0.00063 $), only the light rays with initial angles ($ \Theta $)
 close to $ \pi $ will reach the source circle and produce images. Therefore, we have plotted the right-branch 
 ($ \alpha_R<\Theta<\pi $) and left-branch ($ -\pi<\Theta<\alpha_L $) separately in two different graphs as shown 
in Fig.(\ref{plotRBK}).

\begin{figure}[h]
	\centering
	\begin{subfigure}{.45\textwidth}
		\includegraphics[width=\textwidth]{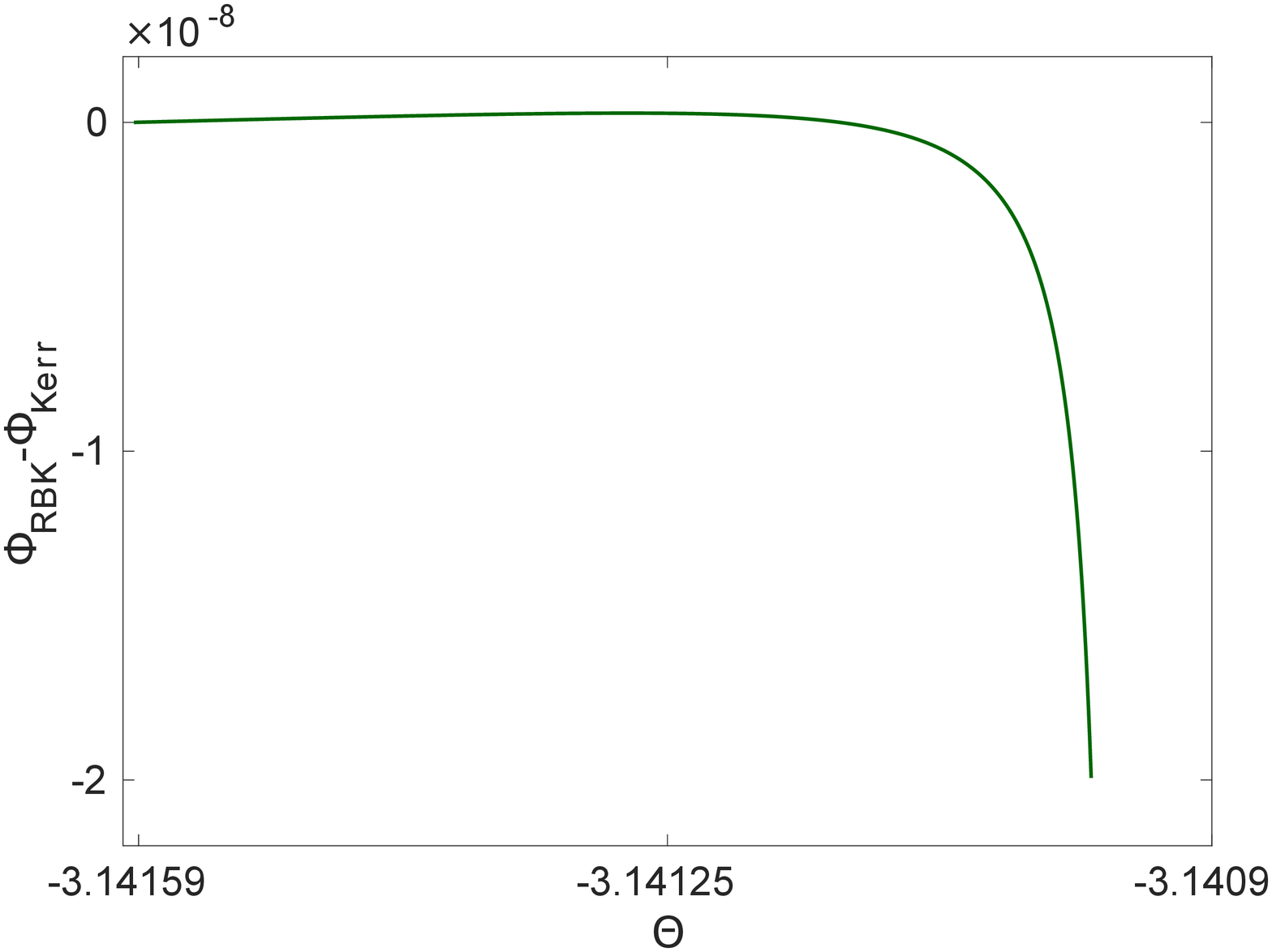}
		\caption{ Left-branch lower portion (real image) }
	\end{subfigure}
	\begin{subfigure}{.45\textwidth}
		\includegraphics[width=\textwidth]{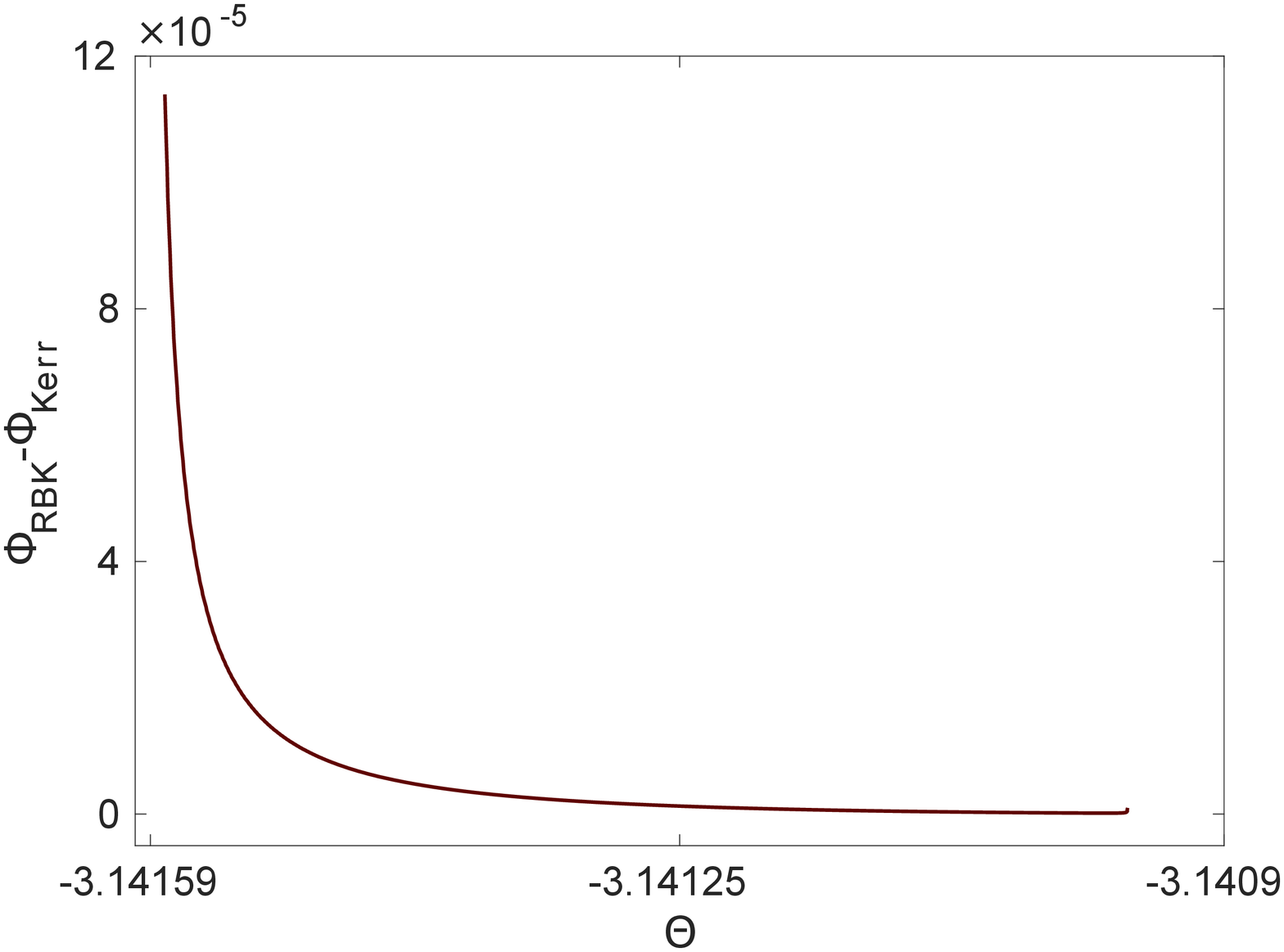}
		\caption{ Left-branch upper portion (hidden image) }
	\end{subfigure}
	\begin{subfigure}{.45\textwidth}
		\includegraphics[width=\textwidth]{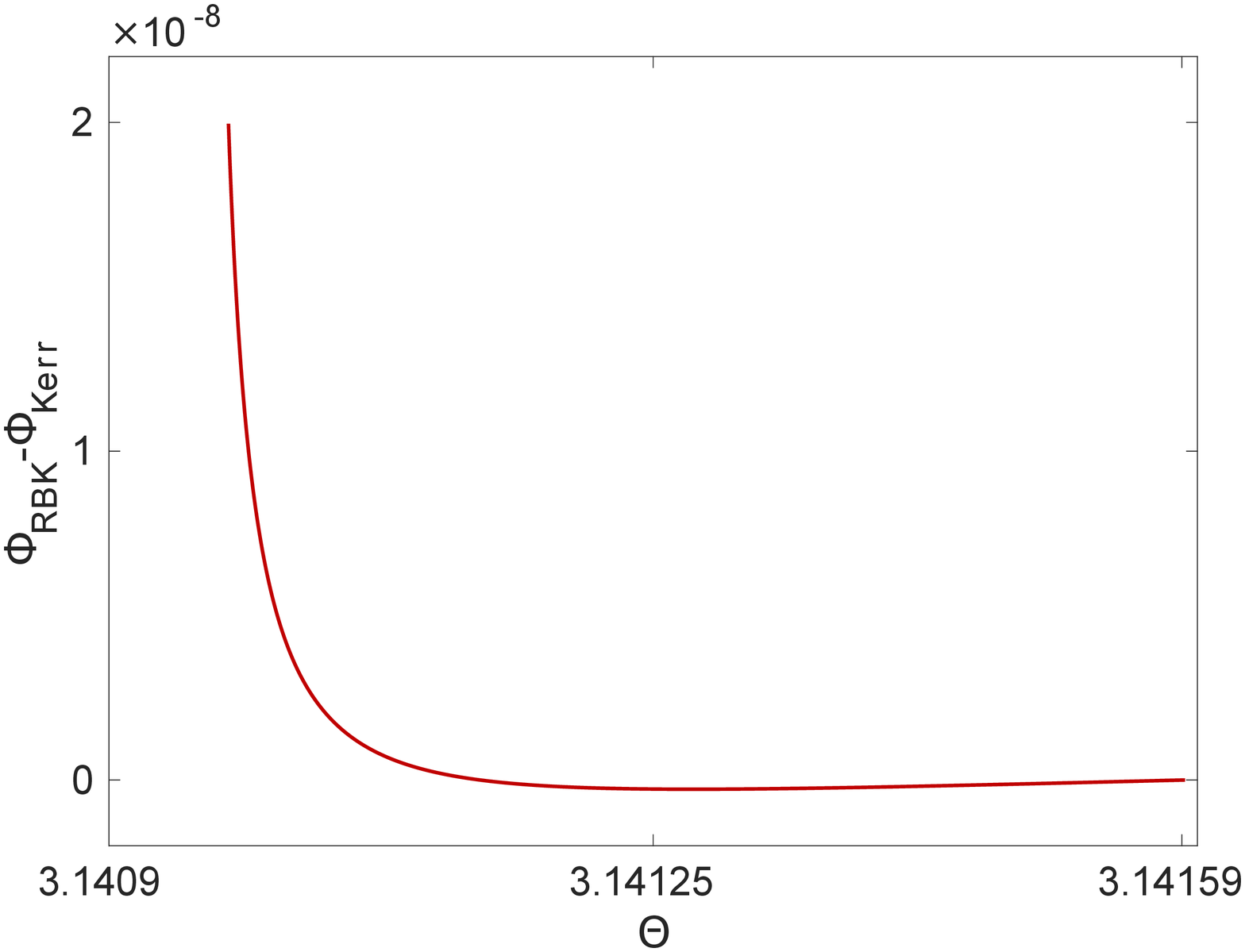}
		\caption{ Right-branch lower portion (real image) }
	\end{subfigure}
	\begin{subfigure}{.45\textwidth}
		\includegraphics[width=\textwidth]{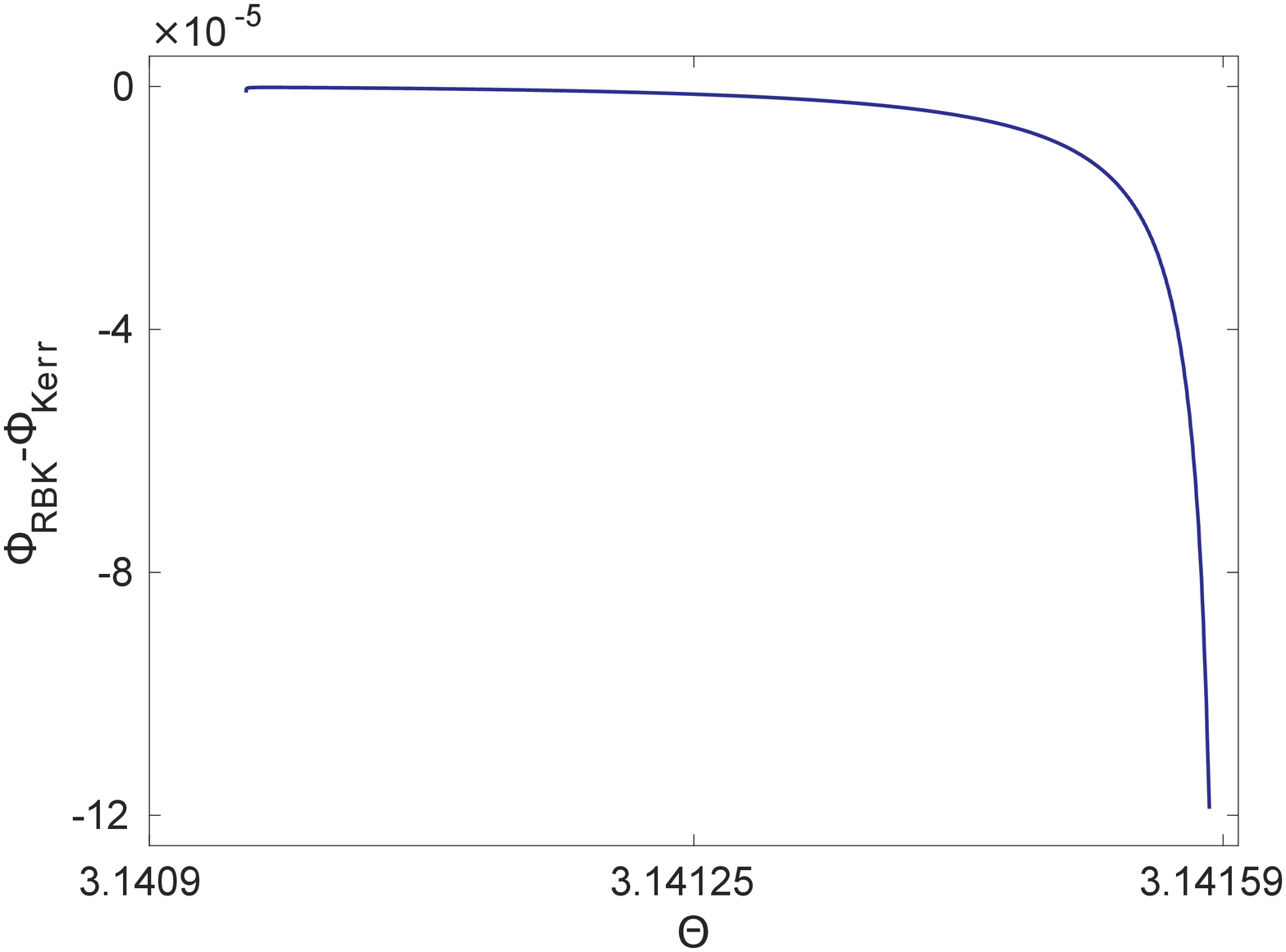}
		\caption{ Right-branch upper portion (hidden image) }
	\end{subfigure}
	\caption{ This figure shows the difference in $ \Phi $ between RBK and Kerr metrics. The left column represents $ \Phi_{RBK}-\Phi_{Kerr} $ 
		for real images, i.e., lower-half portions and the right column represents it for hidden images, i.e. upper-half portions. 
		Plots $ (a) $ and $ (b) $ represent the left-branch, and $ (c) $ and $ (d) $ represent the right-branch. Coloration have been 
		followed as described before. }
	\label{RKBdifference}
\end{figure}

Although the above plots represent characteristic lensing behavior of RBK space-time with data taken from Sombrero galaxy,
 it will be more interesting to calculate the difference in $ \Phi $, i.e., $ \Delta \Phi~(=\Phi_{RBK}-\Phi_{Kerr}) $ for a 
 given value of $ \Theta $, when light rays travel both in RBK and Kerr space-times, and when they travel within 
 Kerr space-time only\footnote{This Kerr metric may be thought of as modeling the central black hole in the galaxy.}, 
 with no matter. This difference will then tell us the contribution in lensing coming from the matter forming the galaxy. 
 Fig.(\ref{RKBdifference}) represents a number of plots showing this difference, $ \Delta \Phi $.

The first and second plots (Fig.(\ref{RKBdifference}a and \ref{RKBdifference}b)) show $ \Delta \Phi (=\Phi_{RBK}-\Phi_{Kerr}) 
$ for the left-branch of $ \Phi-\Theta $ lensing plot. Here, $ (a) $ shows the difference for the lower-half portion of the left-branch, 
which represents $ \Delta \Phi $ for real images, and $ (b) $ shows the difference for the upper-half portion of the same left-branch 
representing hidden images. Similarly, $ (c) $ and $ (d) $ show the corresponding differences for the lower-half and 
upper-half portions of the right-branch respectively. The order of magnitude of $ \Delta \Phi $ for the hidden images are 
found to be larger than the corresponding order of magnitude for the real images (although we note the caveat pointed out earlier
regarding the validity of the RBK metric near the photon sphere). 
The reason for this is that light rays have to travel longer path within the RBK space-time to produce hidden images than 
to produce real images, and so the effect coming from RBK space-time becomes stronger. For the real image plots $ (a) $ 
and $ (c) $, it is found that the difference $ \Delta \Phi $ is almost zero near $ \Theta \simeq \pi $, and start increasing 
rapidly as $ \Theta $ approaches its minimum value for image formation ($ \alpha_R $ or $ \alpha_L$). This happens due to the 
fact that light rays have to travel less distance in going from $ R $ to $ r_S $ in RBK space-time when $ \Theta $ is close to $ \pi $, 
and they travel more distance in RBK space-time, as they are able to bend more, when $ \Theta $ is close to its minimum 
value. Moreover, since we are comparing $ \Phi $ of RBK and Kerr space-times, we also need to incorporate the fact that 
the positions of $ \alpha_R$ (or $ \alpha_L $) are different in the two space-times and this difference also strongly contributes 
in the increment of $ \Delta \Phi $. Although $ \Phi_{RBK}-\Phi_{Kerr} $ increases near $ \alpha_R $ or $ \alpha_L$, 
it never becomes infinite. The difference attains a maximum finite value at $ \alpha_R $ or $ \alpha_L$. 

On the other hand, the opposite happens for hidden images, i.e., $ \Delta \Phi $ increases rapidly near $ \Theta=\pi $ 
and remains almost zero before it, as shown in plots $ (b) $ and $ (d) $. The reason is the same as before, i.e., light rays, 
after crossing the source circle once, travel more path inside the source circle, before crossing it the second time to 
produce hidden images. But the important difference with the real images is that 
$ \Delta \Phi $, in this case, can become large. As the value of $ \Theta $ is very close to $ \pi $, the light 
rays hit the photon circle of RBK space-time which resides very close to its center. As a result, the bending angle of light 
becomes large, so that it makes $ \Phi_{RBK}-\Phi_{Kerr} $ to be large.

In experimental data on lensing, if differences from modeling the central black hole in a galaxy by a Kerr metric are observed, 
then this could constrain the nature of dark matter in these galaxies by the method above. Although it might be premature to 
make any definitive comment on this, the matter certainly deserves further study. 

\subsection{Strong Lensing in RBSTs}

\begin{figure}[H]
	\centering
	\begin{subfigure}{.4\textwidth}
		\includegraphics[width=\textwidth]{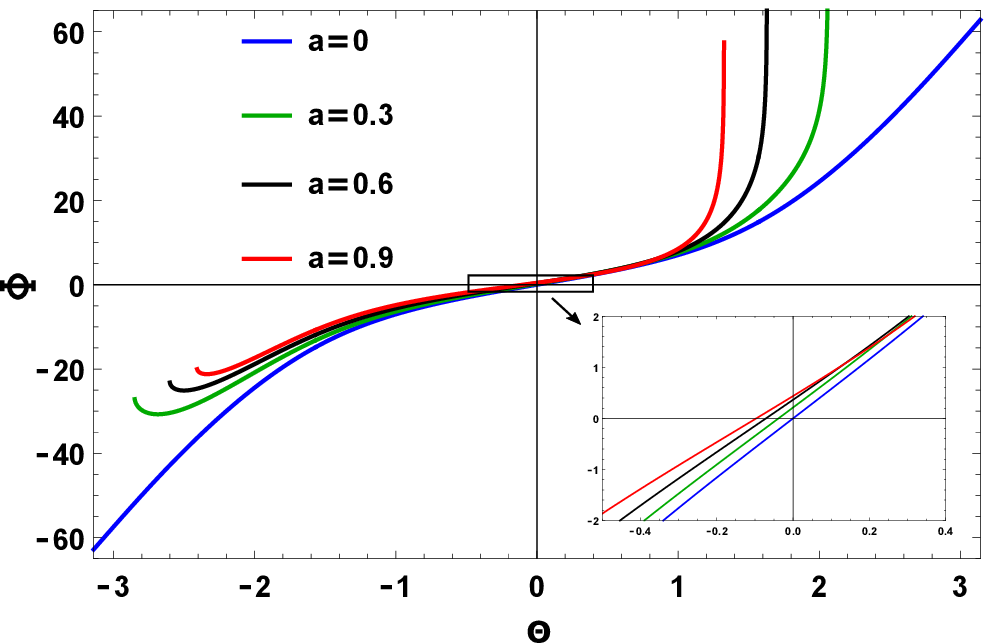}
		\caption{ $ \beta=0.1, ~ r_O = 10, ~ r_S = 20 $ }
	\end{subfigure}
	\begin{subfigure}{.4\textwidth}
		\includegraphics[width=\textwidth]{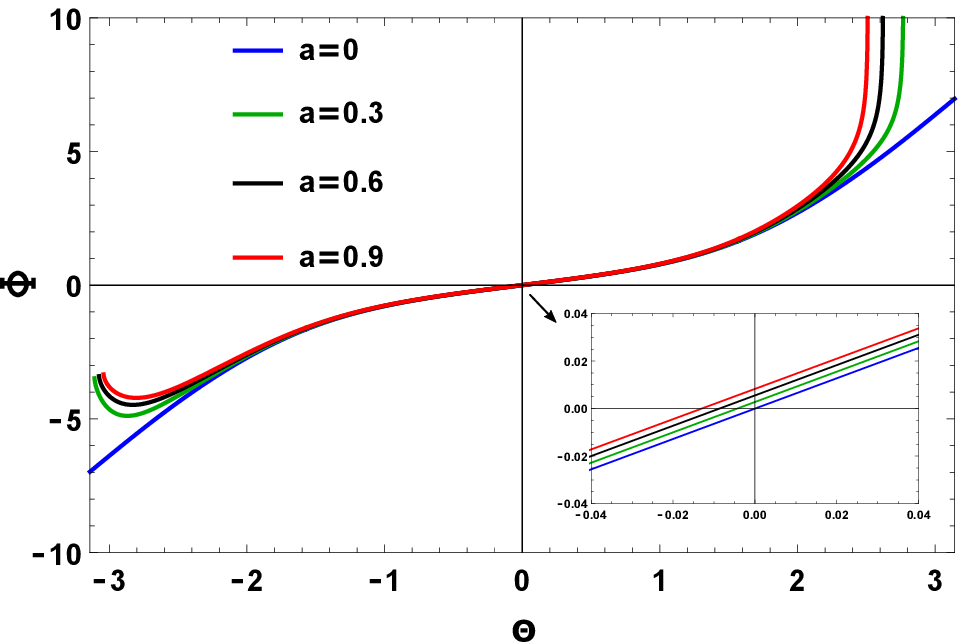}
		\caption{ $ \beta=0.9, ~ r_O = 10, ~ r_S = 20 $ }
	\end{subfigure}
	\begin{subfigure}{.4\textwidth}
		\includegraphics[width=\textwidth]{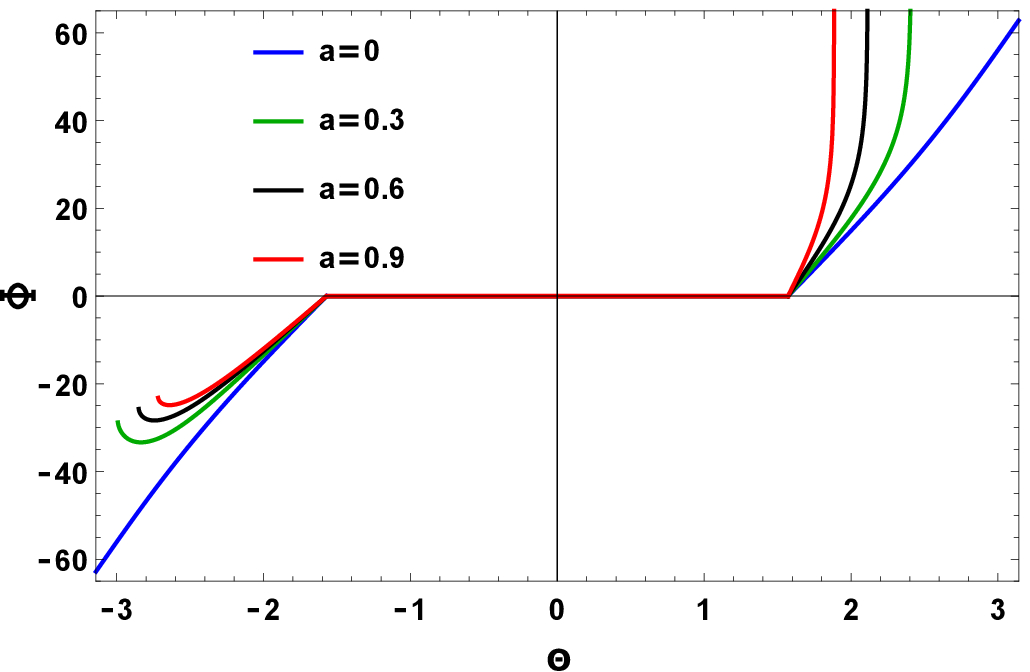}
		\caption{ $ \beta=0.1, ~ r_O = r_S = 20 $ }
	\end{subfigure}
	\begin{subfigure}{.4\textwidth}
		\includegraphics[width=\textwidth]{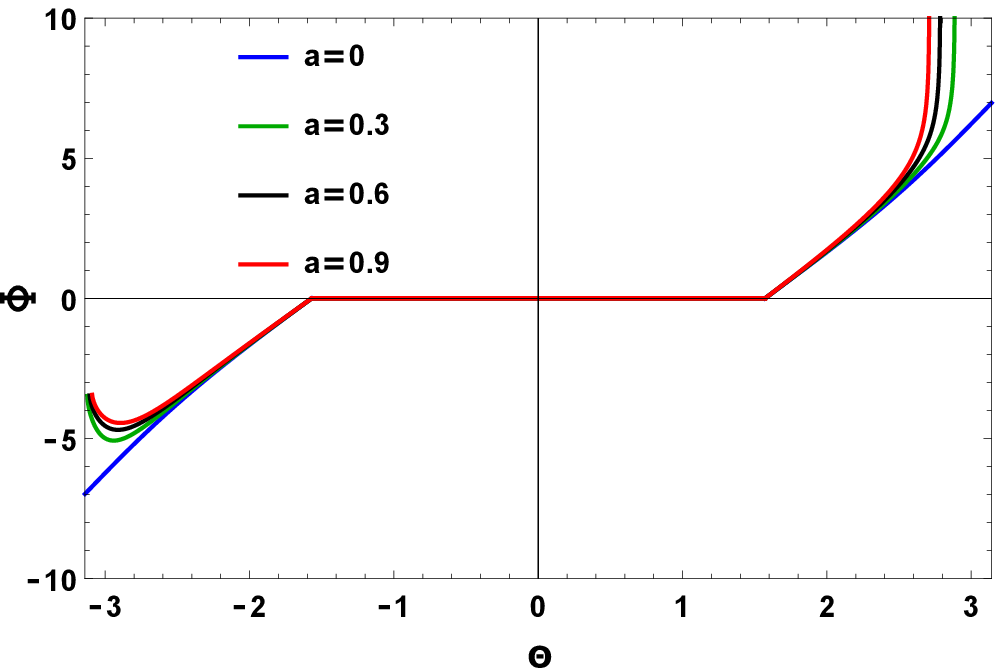}
		\caption{$ \beta=0.9, ~ r_O = r_S = 20 $ }
	\end{subfigure}
	\begin{subfigure}{.4\textwidth}
		\includegraphics[width=\textwidth]{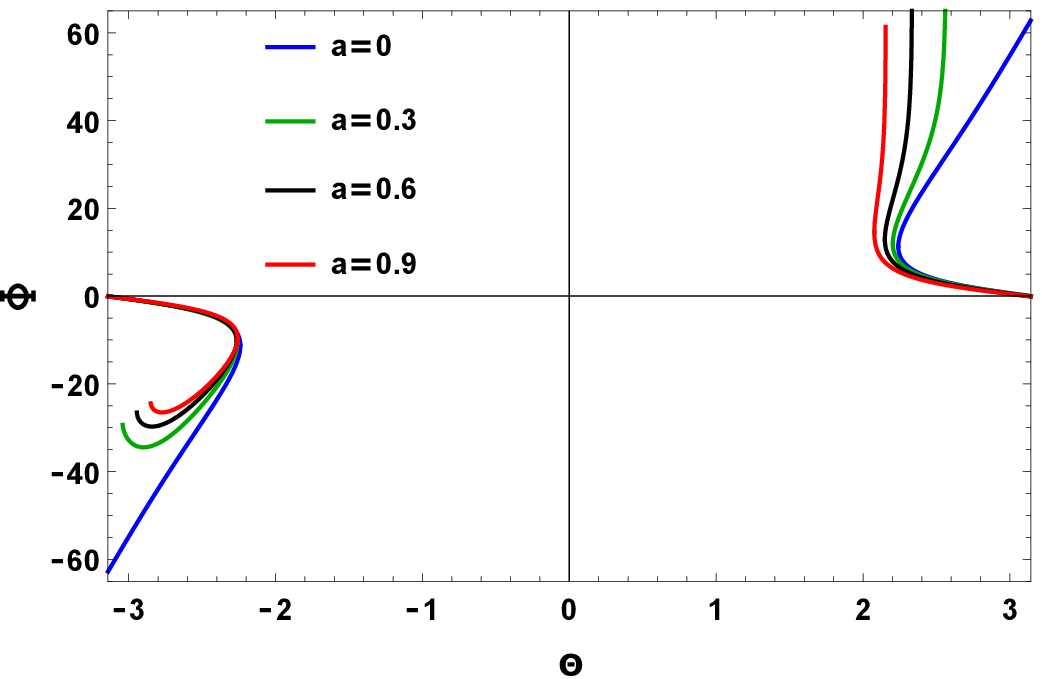}
		\caption{ $ \beta=0.1, ~ r_O = 30, ~ r_S = 20 $ }
	\end{subfigure}
	\begin{subfigure}{.4\textwidth}
		\includegraphics[width=\textwidth]{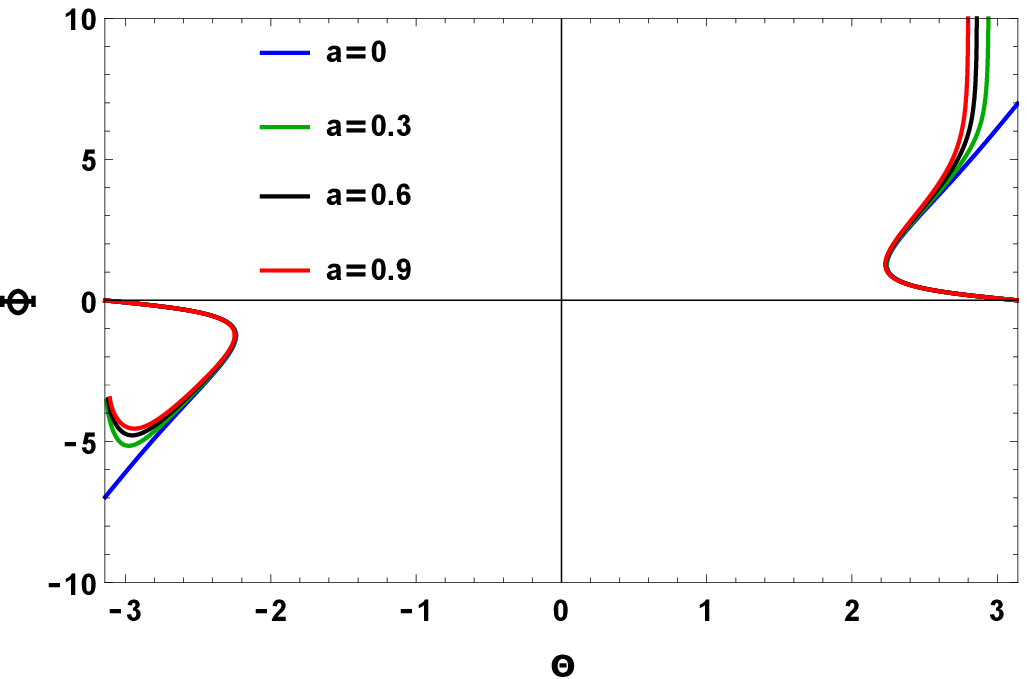}
		\caption{ $ \beta=0.9, ~ r_O = 30, ~ r_S = 20 $ }
	\end{subfigure}
	\caption{$ \Phi $ - $ \Theta $ plots for RBSTs for different values of the rotation parameter, $ a $. }
	\label{fig3}
	
\end{figure}

Now, we study strong lensing in RBST. Different components of the metric of RBST on the equatorial plane are 
given in Eq.(\ref{a13}) of subsection 2.1. 
As before, we have used numerical integral techniques to obtain the lensing plots. We have five different parameters 
in our problem, whose values we need to specify to perform the numerical integration. These are
the galactic length scale $r_h$, radius of the source circle $r_S$, the radius of the observer's circle $r_O$,
the rotation parameter $a$ and $\beta$. 
Fig.(\ref{fig3}) shows a number of plots of the source angular position $ (\Phi) $ as function of image angular position 
$ (\Theta) $ in the observer's sky for different values of the above mentioned parameters. 
For specific values of $ r_O $, $ r_S $ 
and $ \beta $, four different values of $ a $ are considered: $ a = 0.0 $, $ a = 0.3 $, $ a = 0.6 $ and $ a = 0.9 $. Again, for given 
values of $ r_O $ and $ r_S $, two different values of $ \beta $ are considered: $ \beta=0.1 $ (left column) and $ \beta=0.9 $ 
(right column). Plots  (a) and (b)  correspond to $ r_O = 10 $ and $ r_S = 20 $ $ (r_S > r_O ~ \text{case}) $; plots (c) and (d) 
correspond to both $ r_O = 20 $ and $ r_S = 20 $ $ (r_S = r_O ~ \text{case}) $, and pots (e) and (f) correspond to $ r_O = 30 $ 
and $ r_S = 20 $ $ (r_S < r_O ~ \text{case}) $. The value of $ r_h $ is chosen to be, $ r_h = 100 $, in all the plots of Fig.(\ref{fig3}).

For the $ r_S>r_O $ case, the lines pass through the origin when $ a=0 $, i.e., without rotation. This is an artifact of 
the spherical symmetry of static space-times. Introduction of rotation breaks this spherical symmetry and the lines do not
pass through origin. With increasing $ a $, the symmetry breaking becomes stronger and the lines move away from the 
origin, as can be seen from the zoomed in versions in Figs.(\ref{fig3}a and \ref{fig3}b). 

Another important feature is 
that the $ a=0 $ line is finite for the complete range of $ \Theta ~ \in ~ [-\pi , \pi] $. But with increase of rotation, the lines 
start diverging to infinity at lesser values of $ \Theta $ in the range $ 0 < \Theta < \pi $. On the other extreme of 
$ \Theta $ ($ -\pi < \Theta < 0 $), they show an interesting possibility of having two different images of a single source, 
i.e., two values of $ \Theta $ for a single $ \Phi $. First, we need to understand what this divergence of $ \Phi $ mean physically. 
From the ray diagram of lensing shown in Fig.(\ref{fig2}), past oriented light rays are going from the observer's circle to the 
source circle. When a ray reaches the source point after completing a number of turns around the observer circle, 
the value of $ \Phi $ becomes very large for that ray. 

Now, if there exists a photon circle in between $ r_O $ and $ r_S $, 
there will be a maximum value of $ \Theta $ (say, $ \Theta_{max} $) for which the corresponding light ray 
(emerging from the observer with angle $ \Theta_{max} $) will touch the photon circle tangentially and will start 
moving on the photon circle. It will never reach the source circle. For this particular ray, the bending of light is infinite 
and so does $ \Phi $. The light rays emerging form the observer with larger angles than $ \Theta_{max} $ will 
curl into the singularity at $ r=0 $. So the lensing equation will not have any real solution for these type of rays. 
This explains the divergence of the curves in the right hand side of the plots where they reach the photon circle. 
It is evident that faster the rotation parameter $a$, smaller is the value of $ \Theta_{max} $ or in other words, 
the effect of bending is stronger.

On the other hand, since RBST is strongly naked singular for retrograde rays, light rays with negative $ \Theta $ 
will never encounter any photon circle in their paths and correspondingly $ \Phi $ will never diverge. From the left 
side portions of the plots we see that, the curves indeed remain finite upto some values of $ \Theta $ after w
hich there is no real solution of the integral expression of lensing, Eq.(\ref{a6}). Another interesting behavior of the plots 
in this negative $ \Theta $ region $ (-\pi<\Theta<0) $ is observed. For non-zero $ a $, the curves initially go downwards, 
reach minima and then move upwards. This means that two different light rays originating with two different angles 
from the observer's location can meet the source circle at the same source point. 
In other words, one source can have two images located at two different positions in the observer's sky. 

Another point to note here is that both the images are located within the range $ -\pi<\Theta<-\frac{\pi}{2} $. 
If we again go back to Fig.(\ref{fig2}), we see that when $ \Theta $ becomes larger than $ \frac{\pi}{2} $, the corresponding 
ray enters into the $ r=r_O $ circle; goes inside, reaches a minimum radius (say, $\rho$) and then comes out of the 
observer's circle reaching the source location finally. In this situation, $ \dot{r} $ changes sign in going from $ r_O $ to $ r_S $. 
It is negative from $ r_O $ to $ \rho $; zero at $ \rho $ and positive from $ \rho $ to $ r_S $. So we need to replace 
Eq.(\ref{a6}) by a piece-wise integration: from $ r_O $ to $ \rho $ and from $ \rho $ to $ r_S $, as stated earlier. 
Since there is no photon circle for retrograde rays, it is possible that the rays with $ \Theta>\frac{\pi}{2} $ may 
come very close to the singularity where the effects of both curvature and rotation are extremely large. 
Due to these stronger effects, light rays are bending in such a manner that they are producing two different images 
for a single source. Once again, this is happening solely due to the strongly naked nature of the singularity. 
Therefore, the phenomena of bi-image formation can be attributed to the novelty of SNS and can serve as an 
observational tool for detection of SNS, if they exist at all. This conclusion has also been reached for lensing in the 
rotating Janis-Newman-Winicour naked singularity in \cite{Naked}.

The $ r_S = r_O $ case requires special attention here. Equations (\ref{a4}) and (\ref{a6}) become identically 
zero as $ r_S=r_O $ makes both the upper and lower limits of the integral to be equal. So we cannot use the 
integral lens equations directly in this case. From Fig.(\ref{fig2}), it can be clearly seen that, for $ r_S=r_O $,
 $ \dot{r} $ is always non-zero (+ve) when the rays are confined in the range $ -\frac{\pi}{2} < \Theta < \frac{\pi}{2} $. 
 The rays leave the observer circle (which, in this case, happens to be the source circle also) and never 
 return to meet the source circle. Therefore, the value of $ \Phi $ remains zero throughout this range, as can 
 be seen from Figs.(\ref{fig3}c and \ref{fig3}d). Non-trivial lensing behavior is seen only for $ |\Theta|> \frac{\pi}{2} $ 
 and in this range, the nature of the plots are similar to the $ r_S>r_O $ case.

The technical details of the remaining $ r_S < r_O $ case have already been discussed in the previous subsection. 
As stated before, light rays produce hidden images for this case, which can be seen in Figs.(\ref{fig3}e and \ref{fig3}f). 
The reason for the divergence of direct rays with non-zero $ a $ in the right-branches are the same as before.


\section{Discussions and Conclusions}
\label{sec-5}

We will now summarize the main results of this paper. 

\begin{itemize}
	\item 
	We have obtained possible rotating galactic space-times by applying the Newman-Janis algorithm to static galactic seed metrics. We have
	considered two seed metrics here, the Bharadwaj-Kar metric obtained from phenomenological grounds, and Bertrand space-times derived
	from the assumption that there exists stable circular orbits at all values of the radial coordinate. The first solution was matched on a time-like
	hypersurface with a Kerr metric. The second solution is theoretically 
	interesting, and we have provided a rotating solution of Einstein's equations that generalizes Bertrand's theorem in classical mechanics and
	admits stable circular orbits at all radii. In both	cases, the weak energy conditions have been checked. 
	\item 
	We have obtained a rotating generalization of Perlick's strong lensing formalism that allow for the source and observer to be at finite
	distances, and further does not assume an asymptotically flat metric. 
	\item
	Strong gravitational lensing has been studied using the above formalism for RBK and RBST solutions. In the former, we have pointed out 
	the differences that might arise compared to lensing from Kerr black hole. In the latter, the possibility of bi-image formation is discussed.  
\end{itemize}

   We now point out a few caveats in our analysis. First, as pointed out in the main text, the computation of the hidden images in the RBK metric
   assumed that this is valid up to the photon circle. As we have said, this might not generically be the case, and the issue deserves
   further analysis. Next, we have not performed a stability analysis of the RBST. Note here that
   the RBK space-times are valid away from galactic centers. RBST on the other hand is theoretically valid in $0<r<\infty$. In this context,
   we are aware of the work of \cite{Suneeta} (see also references therein) where stability analysis of a class of static naked singularities was
   considered and it was shown that such singularities might be stable under some types of metric perturbations. BSTs and their
   rotating generalizations seem more difficult to handle, and we hope to report on this aspect elsewhere. 

\newpage
\begin{center}
{\bf Acknowledgements}
\end{center}
We sincerely thank Sayan Kar for an useful email correspondence. 

\bibliographystyle{utphys}
\bibliography{References.bib}

\end{document}